\pdfoutput=1

\documentclass[natbib]{svjour3}                     
\smartqed  
\usepackage{graphicx}
 \usepackage{mathptmx}      
 \usepackage{amsmath}
 \usepackage{rotating}
 \usepackage{verbatim}
 \usepackage[misc]{ifsym} 
 \usepackage{floatrow}
 \usepackage[table]{xcolor}
\newcommand{\vlos}{{V_{\rm los}}}
\begin{document}
 
\title{Helioseismology with {\it Solar Orbiter}
}


\author{Bj\"orn~L\"optien \and Aaron~C.~Birch \and Laurent~Gizon \and Jesper~Schou \and Thierry~Appourchaux \and Juli{\'a}n~Blanco~Rodr{\'i}guez \and Paul~S.~Cally \and Carlos~Dominguez-Tagle \and Achim~Gandorfer \and Frank~Hill \and Johann~Hirzberger \and Philip~H.~Scherrer \and Sami~K.~Solanki}


\institute{B. L\"optien \and L. Gizon (\Letter)  \at
Institut f\"ur Astrophysik, Georg-August Universit\"at G\"ottingen, 37077 G\"ottingen, Germany\\
\email{gizon@mps.mpg.de}
\and A.~C. Birch \and L. Gizon \and J. Schou \and A. Gandorfer \and J. Hirzberger \and S.~K. Solanki \at
Max-Planck-Institut f\"ur Sonnensystemforschung, Justus-von-Liebig-Weg 3, 37077 G\"ottingen, Germany
\and T. Appourchaux \and C. Dominguez-Tagle \at
Institut d'Astrophysique Spatiale, CNRS - Universit\'{e} Paris Sud, 91405 Orsay, France
\and J. Blanco Rodr{\'i}guez \at
GACE/IPL, Universidad de Valencia, 46980 Valencia, Spain
\and P.~S. Cally \at
Monash Centre for Astrophysics and School of Mathematical Sciences, Monash University, Clayton, Victoria 3800, Australia
\and F. Hill \at
National Solar Observatory, Tucson, Arizona 85719, USA
\and P.~H. Scherrer \at
Hansen Experimental Physics Laboratory, Stanford University, Stanford, CA 94305-4085, USA}

\date{Received: date / Accepted: date}

\maketitle

\begin{abstract}
The {\it Solar Orbiter} mission, to be launched in July 2017, will carry a suite of remote sensing and in-situ instruments, including the {\it Polarimetric and Helioseismic Imager} (PHI). PHI will deliver high-cadence images of the Sun in intensity and Doppler velocity suitable for carrying out novel helioseismic studies. The orbit of the {\it Solar Orbiter} spacecraft will reach a solar latitude of up to $21^\circ$ (up to $34^\circ$ by the end of the extended mission) and thus will enable the first local helioseismology studies of the polar regions. Here we consider an array of science objectives to be addressed by helioseismology within the baseline telemetry allocation (51 Gbit per orbit, current baseline) and within the science observing windows (baseline $3\times 10$ days per orbit). A particularly important objective is the measurement of large-scale flows at high latitudes (rotation and meridional flow), which are largely unknown but play an important role in flux transport dynamos. For both 
helioseismology and feature tracking methods convection is a source of noise in the measurement of longitudinally averaged large-scale flows, which decreases as $T^{-1/2}$ where $T$ is the total duration of the observations. Therefore, the detection of small amplitude signals (e.g., meridional circulation, flows in the deep solar interior) requires long observation times. As an example, one hundred days of observations at lower spatial resolution would provide a noise level of about three~m/s on the meridional flow at 80$^\circ$~latitude. Longer time-series are also needed to study temporal variations with the solar cycle. The full range of Earth-Sun-spacecraft angles provided by the orbit will enable helioseismology from two vantage points by combining PHI with another instrument: stereoscopic helioseismology will allow the study of the deep solar interior and a better understanding of the physics of solar oscillations in both quiet Sun and sunspots. We have used a model of the PHI instrument to study its 
performance for helioseismology applications. As input we used a 6~hr time-series of realistic solar magneto-convection simulation (Stagger code) and the SPINOR radiative transfer code to synthesize the observables. The simulated power spectra of solar oscillations show that the instrument is suitable for helioseismology. In particular, the specified point spread function, image jitter, and photon noise are no obstacle to a successful mission.

\end{abstract}

\clearpage

\setcounter{tocdepth}{5}
\tableofcontents

\clearpage

\section{Introduction: Solar Orbiter} \label{sect:intro}
{\it Solar Orbiter}\footnote{The web page of the mission is located here: http://sci.esa.int/solar-orbiter/} is an ESA M-Class mission and Europe's follow-up to the highly successful SOHO mission of ESA and NASA. It was selected in 2011 with a baseline launch date of July 2017. An important feature of {\it Solar Orbiter} is its orbit; the spacecraft will approach the Sun as closely as 0.28 AU and reach heliographic latitudes of up to 34$^\circ$, which will allow {\it Solar Orbiter} to directly observe the solar poles at a much lower angle than is possible from Earth. The spacecraft will host a suite of in-situ and remote-sensing instruments. Direct measurements within the inner heliosphere can be related directly to observations of the different layers of the solar atmosphere, from the photosphere through the chromosphere, the corona, and the inner heliosphere. This will provide unique opportunities for studying the connection between the Sun and the heliosphere. As described in the {\it Solar Orbiter 
Definition Study Report}~\citep{Redbook} and in~\citet{2013SoPh..285...25M}, the central science goal of the mission is:

\noindent {\it How does the Sun create and control the heliosphere?}

\noindent In order to answer this question, {\it Solar Orbiter} will address four separate science questions:
\begin{itemize}
\item How and where do the solar wind plasma and magnetic field originate in the corona?
\item How do solar transients drive heliospheric variability?
\item How do solar eruptions produce the energetic particle radiation that fills the heliosphere?
\item How does the solar dynamo work and drive the connections between the Sun and the heliosphere?
\end{itemize}

\noindent One key aspect for addressing these science questions, particularly the last one, will be probing the solar interior by using helioseismology. Helioseismology is expected to contribute significantly to the determination of subsurface flows in the upper solar convection zone, especially in the polar regions. This will offer unique opportunities to study the dynamics of plasma flows and the solar dynamo. By combining observations from other instruments in space or on the ground, it will also be possible to experiment with stereoscopic helioseismology to probe the deep solar interior.

\section{Mission Profile} \label{sect:mission}
\subsection{Orbit Design}
\begin{figure}
 \centering
\includegraphics[width=0.8\textwidth]{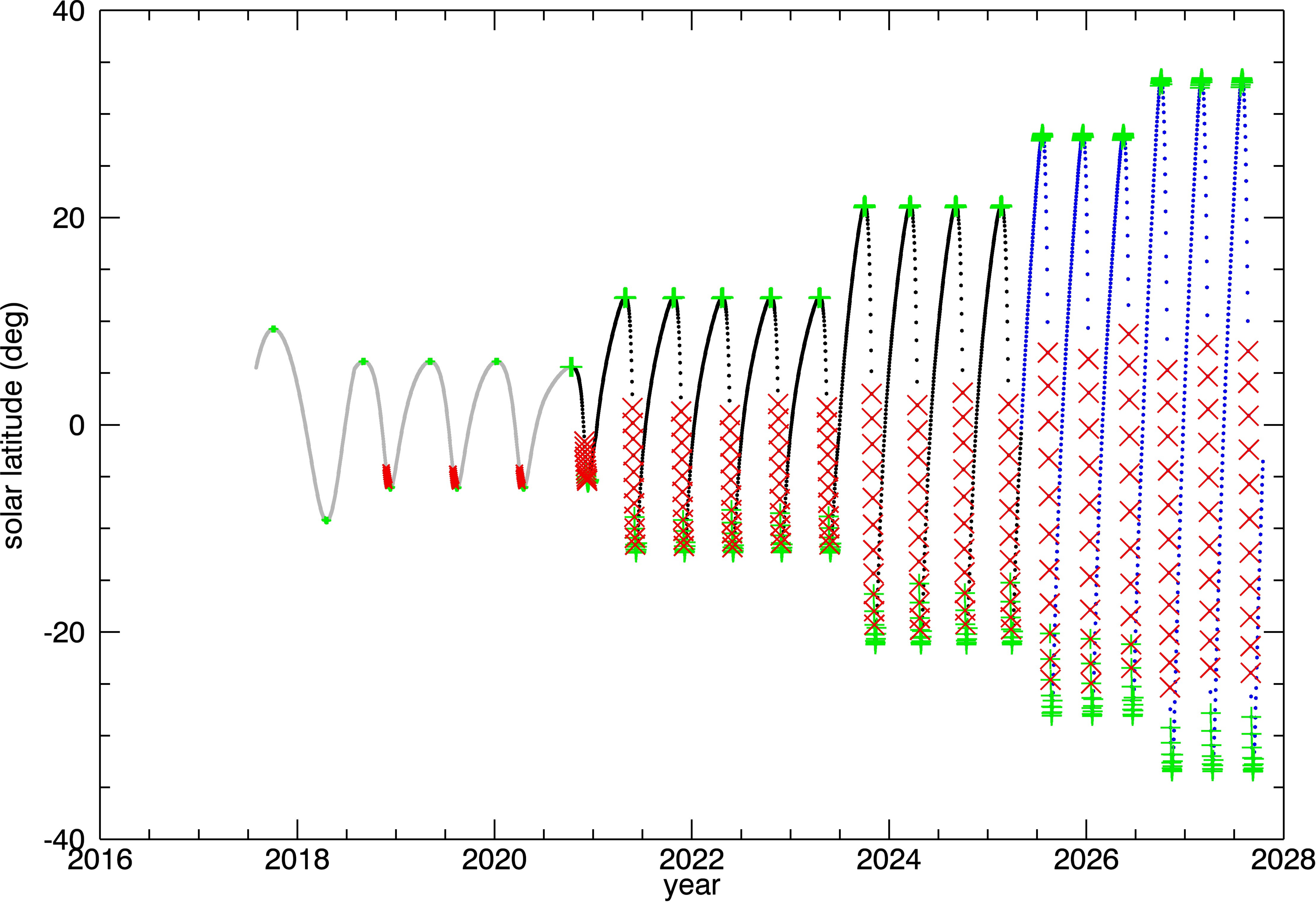}
\includegraphics[width=0.8\textwidth]{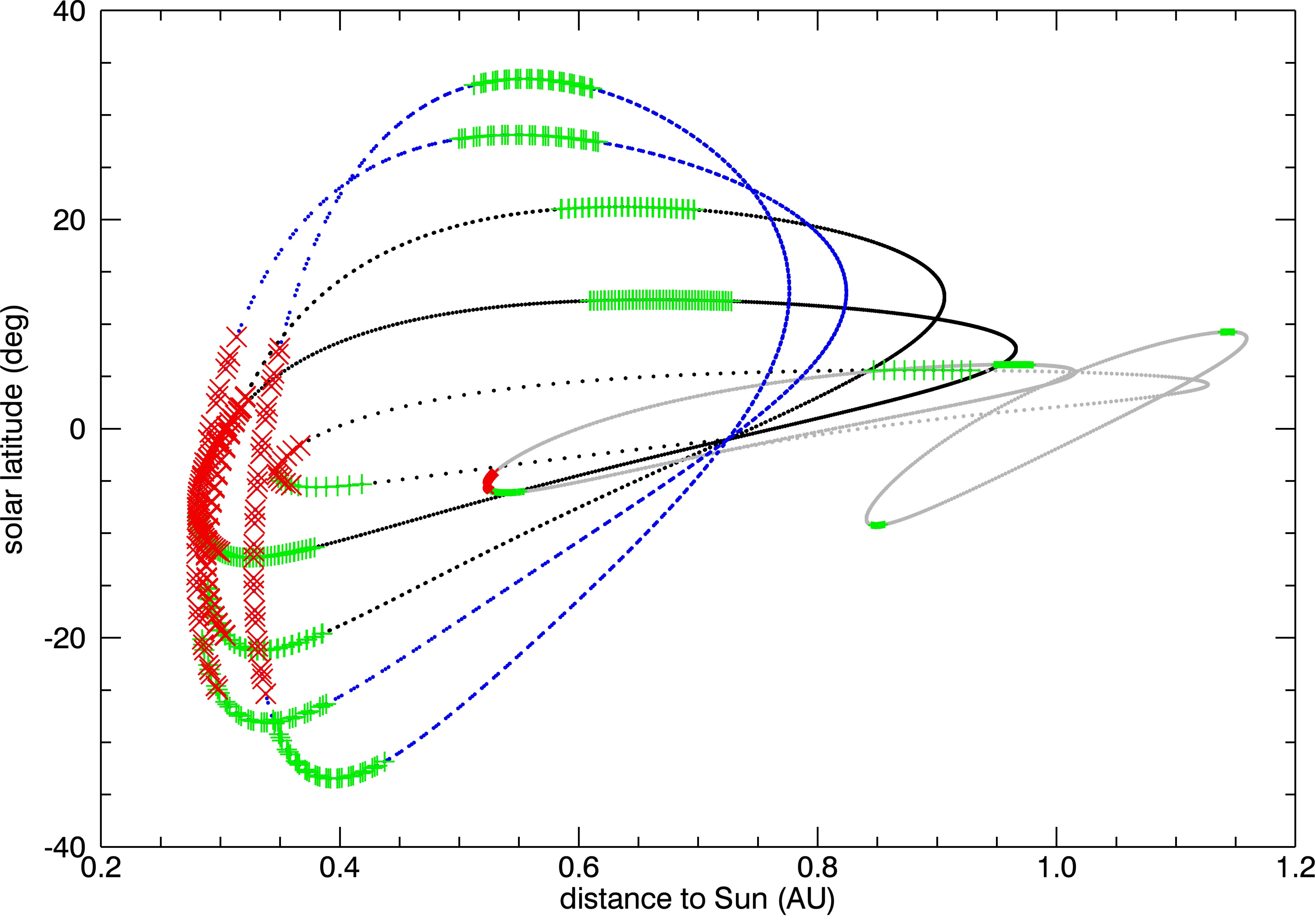}
\includegraphics[width=0.8\textwidth]{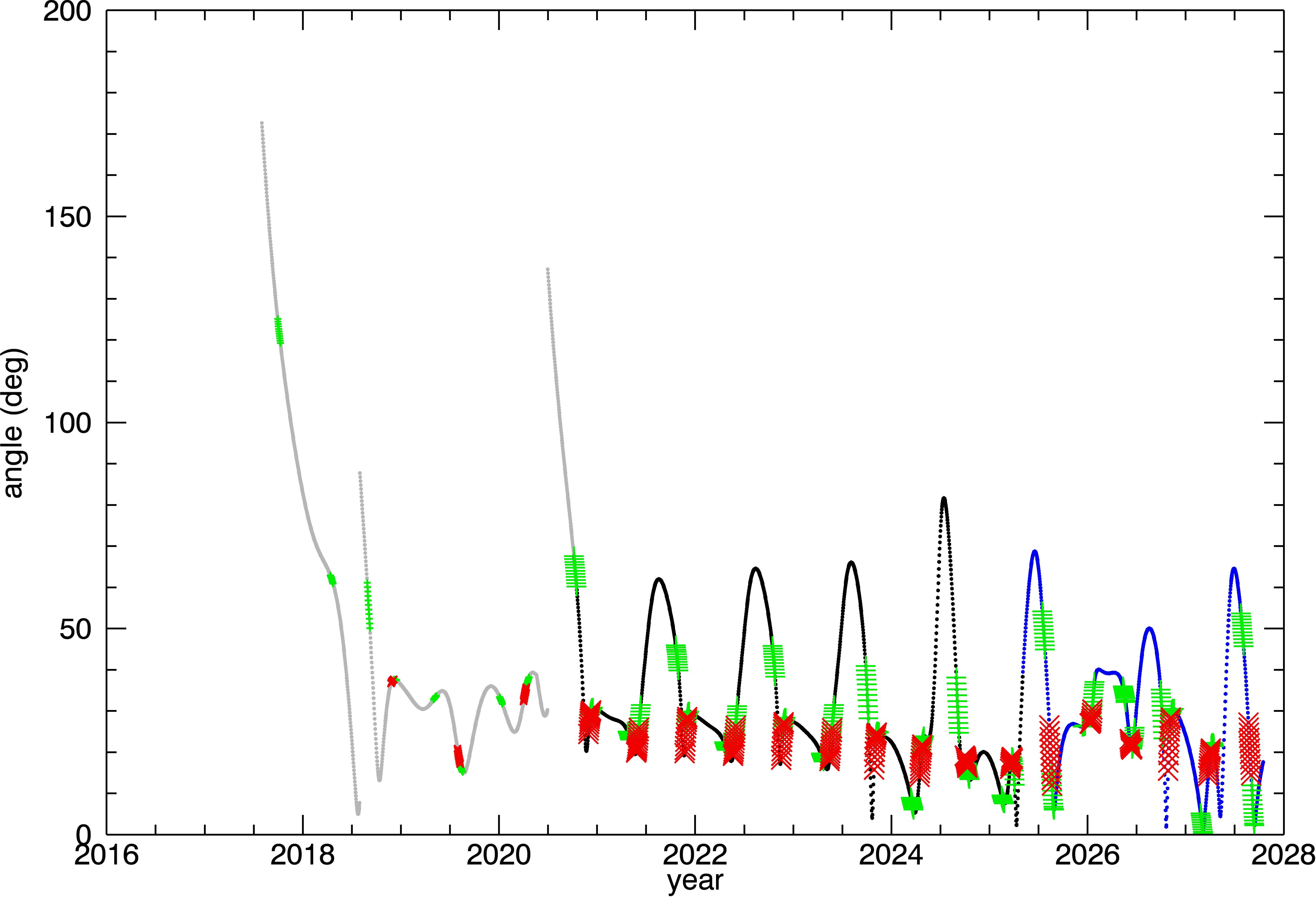}
\caption{Orbit of {\it Solar Orbiter} (July 2017 baseline launch). {\it From top to bottom:} Solar latitude as a function of time, solar latitude as a function of distance to the Sun and the angle Sun-Earth-spacecraft as a function of time. The {\it black dots} show the orbit for the science nominal mission and the {\it gray} and {\it blue dots} denote the cruise phase and the extended mission phase. The current baseline for the remote-sensing instruments is to observe in only three science windows per orbit (at perihelion and maximum northern/southern heliographic latitude). The science windows are given by the {\it red X-symbols} (perihelion) and {\it large green crosses} (maximum latitude). The {\it small symbols} show the perihelion and the maximum latitude during the cruise phase. The spacecraft will reach a maximum solar latitude of $21^\circ$ during the nominal mission and up to $34^\circ$ during the extended mission. The closest perihelion is at 0.28 AU}
\label{fig:orbit2}
\end{figure}

{\it Solar Orbiter} draws its unique capabilities from its special orbit characteristics. 
After separation from the launch vehicle, {\it Solar Orbiter} will start its $3$~year cruise phase. Subject to several Gravity-Assist-Maneuvers (GAMs) at Venus and Earth, the spacecraft will lose orbital energy, which will allow {\it Solar Orbiter} to reduce the distance to the Sun. After a second GAM at Venus, {\it Solar Orbiter} begins its operational phase. From then on its orbit is in resonance with Venus with an initial period of about $180$~days, such that the inclination of the orbital plane with respect to the ecliptical plane can be increased by Venus gravity assists. This particular and unique feature gives {\it Solar Orbiter} access to the high latitude regions of the Sun (see Figure~\ref{fig:orbit2} and Table~\ref{tab:orbit}). The final orbit will be highly eccentric with a perihelion distance of $0.28$~AU. At perihelion, {\it Solar Orbiter} will have a low angular velocity relative to the Sun. This will allow {\it Solar Orbiter} to follow the evolution of surface structures and solar 
features for a longer time than possible from Earth (19 days with a viewing angle between $\pm 75^\circ$). The nominal mission phase lasts for about $4.5$~years. Afterward, the mission could be extended for another $3.5$~years.


 
\begin{table}
\caption{Orbit parameters for a launch in July 2017}
\label{tab:orbit}       
\begin{tabular}{lll}
\hline\noalign{\smallskip}
 & Nominal mission & Extended mission\\
\noalign{\smallskip}\hline\noalign{\smallskip}
Starting date & 2020& 2025\\
Closest distance [AU] & 0.28 & 0.28\\
Max. solar latitude [$^\circ$] & 21 & 34\\
Max. relative velocity [km/s] & 25 & 23\\
\noalign{\smallskip}\hline
\end{tabular}
\end{table}

\subsection{Instrument Suite}
One of the most important aspects  of the {\it Solar Orbiter} mission is the combination of remote observing with in-situ measurements. Further, the high-resolution instruments are all designed to observe the same target region on the solar surface. This will allow relating observations of different atmospheric layers, which will be seen using the different instruments. The instruments of {\it Solar Orbiter} are described in detail in the {\it Solar Orbiter Definition Study Report}~\citep{Redbook}. They can be grouped in three major packages, each consisting of several instruments:
\begin{itemize}
\item Field Package:
{\it Radio and Plasma Waves Instrument} (RPW) and {\it Magnetometer} (MAG).
\item Particle Package:
{\it Energetic Particle Detector} (EPD) and {\it Solar Wind Plasma Analyzer} (SWA).
\item Solar remote sensing instrumentation:
{\it Polarimetric and Helioseismic Imager} (PHI), {\it Extreme Ultraviolet Imager} (EUI), {\it Multi Element Telescope for Imaging and Spectroscopy} (METIS), {\it Solar Orbiter Heliospheric Imager} (SoloHI), {\it Spectral Imaging of the Coronal Environment} (SPICE) and {\it Spectrometer/Telescope for Imaging X-Rays} (STIX).
\end{itemize}
The {\it Polarimetric and Helioseismic Imager} (PHI) will provide the vector magnetic field and line-of-sight (LOS) velocity on the visible solar surface at high spatial resolution, thanks to the close-by observing conditions during the perihelion passages. The magnetic field anchored at the solar surface produces most of the structures and energetic events in the upper solar atmosphere and significantly influences the heliosphere. Extrapolations of the magnetic field observed by PHI into the Sun's upper atmosphere and heliosphere will provide the information needed for other optical and in-situ instruments to analyze and understand the data recorded by them in a proper physical context. In addition, the Dopplergrams or intensity images obtained with PHI will be used to measure solar oscillations and thereby allow probing of the solar interior using helioseismology.


\subsection{PHI: Observables and Operation}

PHI is based on a tunable narrow-band filtergraph which is designed to scan the Fe I $6173$~\AA \ line~\citep[the same as observed by HMI,][]{2012SoPh..275..229S} with a planned cadence of $60$~s. This line forms in a broad range of heights, ranging from about 100 to 400 km above the photosphere~\citep{2011SoPh..271...27F}. Due to telemetry constraints, the instrument will perform onboard inversion for line-of-sight velocity and magnetic field vector. In addition to the inverted quantities the continuum intensity, which will be measured directly by PHI, will be transferred to ground.

\begin{figure}
\begin{minipage}[h!]{\textwidth}
\includegraphics[width=0.5\textwidth]{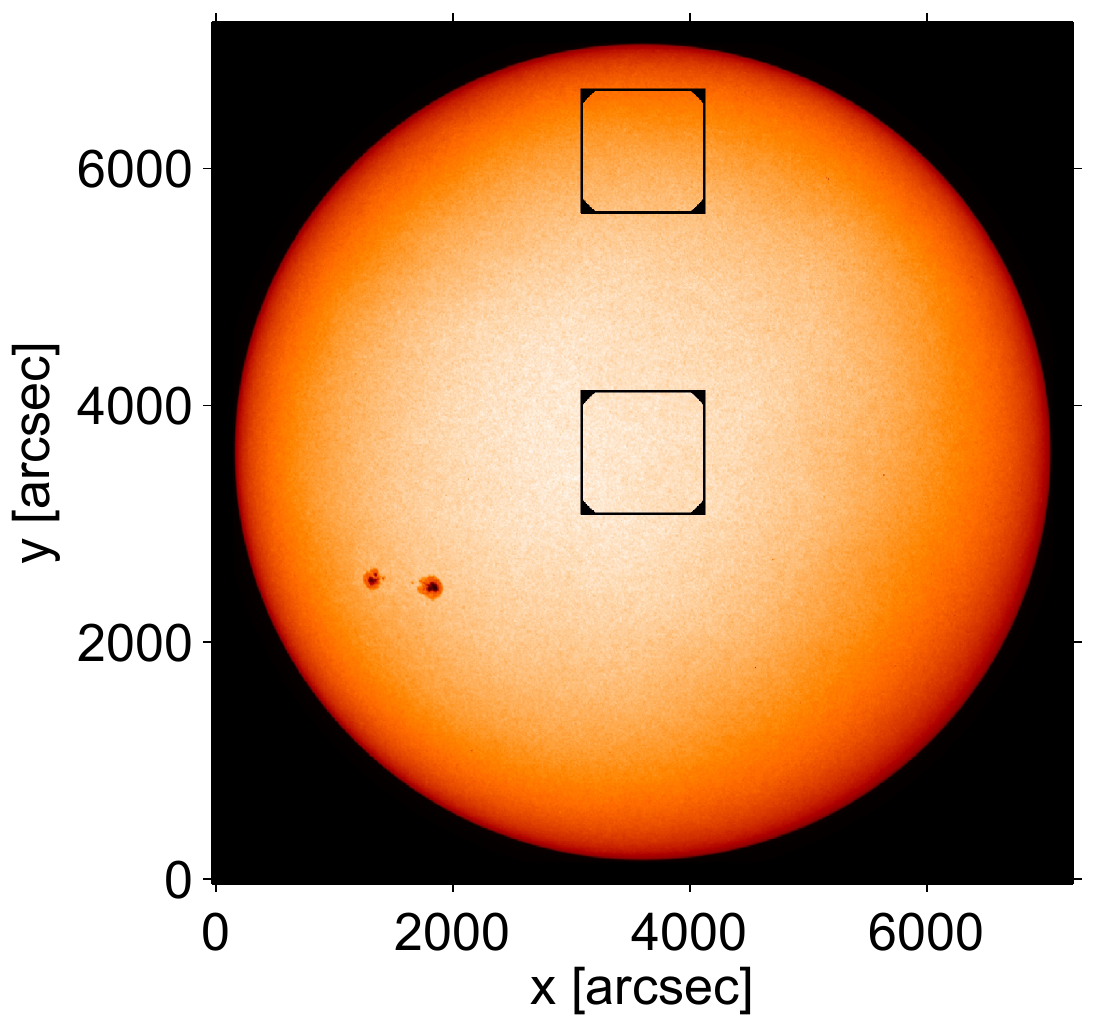}
\includegraphics[width=0.5\textwidth]{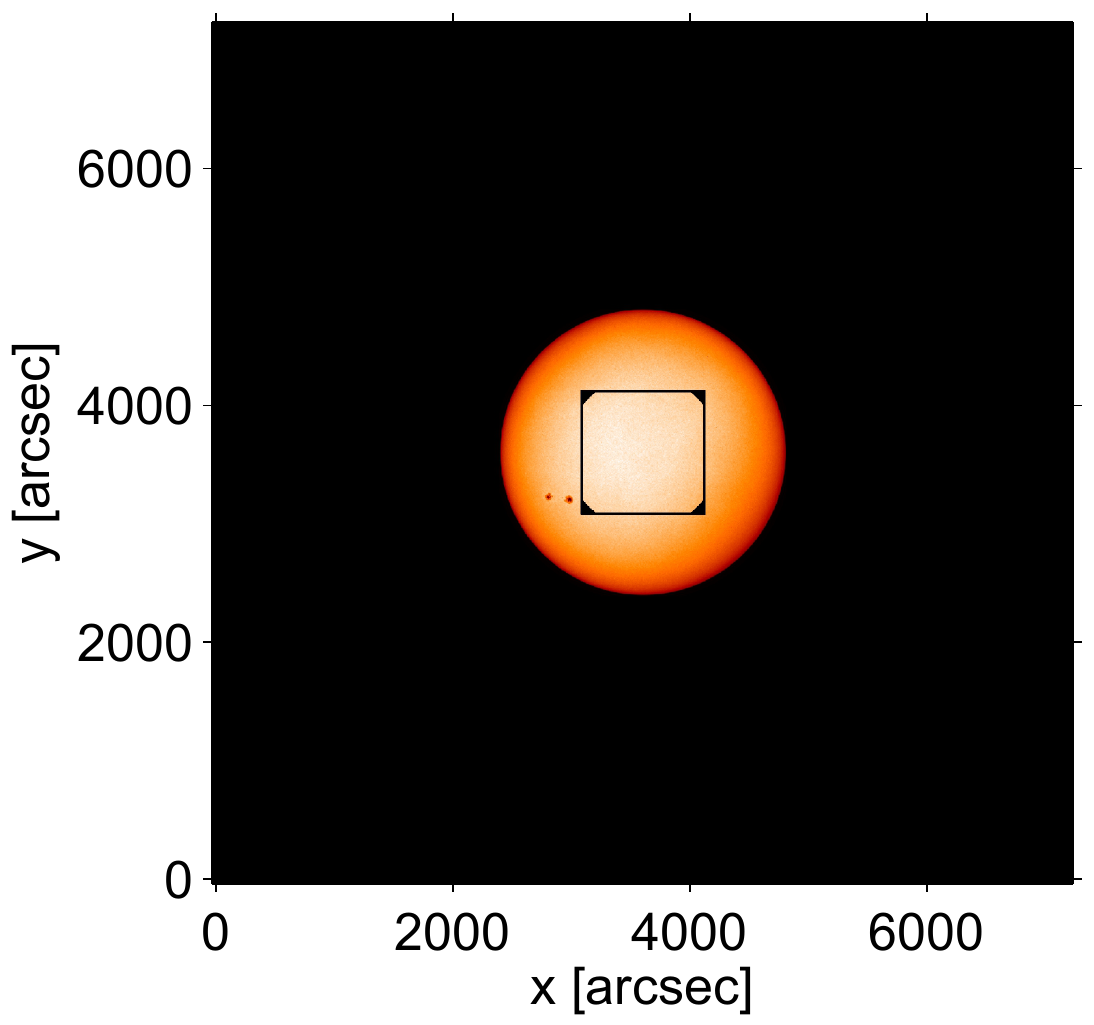}
\end{minipage}
\caption{Field-of-view (FOV) of PHI at $0.28$~AU ({\it left}) and $0.8$~AU ({\it right}) solar distance. The {\it full images} show the FOV of the FDT and the {\it small squares} that of the HRT. Although PHI will not have a pointing mechanism of its own, observations close to the limb with the HRT will be possible by changing the pointing of the spacecraft ({\it upper small square in the image on the left})}
\label{fig:FOV}
\end{figure}

PHI will consist of two telescopes that cannot be used simultaneously, a Full Disk Telescope (FDT) with a field of view of $2^\circ$, sufficient for observing the entire solar disk during the whole orbit with a spatial resolution of about $9''$ ($\sim 1800$~km at disk center at perihelion) and a High Resolution Telescope (HRT) that will image a fraction of the Sun with high spatial resolution. The HRT has a field of view of $16.8'$ and a spatial resolution of about $1''$, corresponding to 200 km at disk center at perihelion. As can be seen in Figure~\ref{fig:FOV}, the FOV and the resolution of the HRT and FDT vary significantly over the course of the orbit.

The current baseline for PHI and the other remote sensing instruments is to observe only during three science windows per orbit (see Figure~\ref{fig:orbit2}), at perihelion and at maximum northern/southern heliographic latitude. Each science window is planned to last for $10$~days. Observations outside of the science windows are in principle possible but they require the spacecraft to provide accurate pointing and thermal stability. Also, operating the remote-sensing instruments might interfere with measurements taken by the in-situ instruments. Helioseismology, however, requires long observing times and is thus a strong argument for increasing the length of the science windows.

PHI is also affected by the low telemetry rate of {\it Solar Orbiter}. PHI has an allocated telemetry rate of $51$~Gbits per science orbit, which corresponds to $\sim 3.3$~kbps on average. This is not sufficient for transmitting continuous observations with full resolution to Earth. Telemetry is thus a constraint for helioseismology and motivates a well-planned observing strategy. PHI will perform extensive onboard processing of the data, consisting of image processing and radiative transfer equation (RTE) inversion. For larger data volumes, additional compression will be applied. Since the same computational resources are used for data acquisition and processing, the processing can only be made when no observations are performed. The instrument will host $4$~Tbits of flash memory, which will allow storage of raw and processed data. Data obtained within the science windows can be processed during the whole orbit and transmitted to Earth whenever telemetry is available. In addition to the inverted 
quantities, raw data will be transferred as well from time to time for testing purposes. When obtaining data for helioseismology only, a full inversion is not required. Instead, a simple algorithm for determining the LOS velocity can be used, as was  successfully done by MDI~\citep{1995SoPh..162..129S}.


\section{Helioseismology: Science Objectives}
\label{sect:science}

The helioseismology science objectives for {\it Solar Orbiter} include measuring differential rotation, torsional oscillations, meridional flow, and convective flows at high latitudes and, also, in combination with another instrument~\citep[e.g., HMI or GONG++, ][]{2003ESASP.517..295H} using stereoscopic helioseismology to probe the deep convection zone and the tachocline. These science objectives are discussed in the {\it Solar Orbiter Definition Study Report}~\citep{Redbook}. Discussions of the science objectives involving helioseismology are also given in~\citet{Yellowbook}, \citet{2001ESASP.493..227G} and~\citet{2007AN....328..362W}.

\begin{table}
\caption{Choices that have to be made in planning an observing scheme for helioseismology}
\label{tab:obs}
\begin{tabular}{p{2.5cm}p{3cm}p{5cm}}
\hline\noalign{\smallskip}
Parameter & Options & Baseline for helioseismology (this paper) \\
\noalign{\smallskip}\hline\noalign{\smallskip}
Observables & Doppler ($\vlos$), $I_{\rm c}$, $B_{\rm los}$ & $\vlos$\\
Cadence, $dt$ & $> 10$ s & 60 s\\
Duration, $T$ & Set by telemetry & 10 days nominal, hopefully much more\\
Telescope & FDT, HRT & Depends on application\\
Spatial sampling & Open & A few samples per min. wavelength \\
Compression & Set by desired p-mode S/N & 5 bit per observable\\
Vantage point & Set by orbit & High inclination to study polar regions; \newline Full range of Earth-Sun-spacecraft angles for stereoscopy\\
\noalign{\smallskip}\hline
\end{tabular}
\end{table}
Here we present an array of science objectives for helioseismology that could be addressed by PHI. We focus on science targets that we expect will go beyond what is possible, or will be possible, with current or future observations from HMI, MDI, GONG++, or BiSON~\citep{1996SoPh..168....1C}. For all of the science objectives presented here we make suggestions for an observing strategy, in all cases we give both the observing time and the number of pixels (see Table~\ref{tab:obs} for a list of parameters). In order to reduce telemetry, we suggest rebinning the data to a given resolution (a certain number of pixels across the solar disk). Most of the time, observations would be performed using the FDT. For some science goals, this resolution cannot be achieved when using the FDT when observing around aphelion. In this case, observations should be performed using the HRT, which will also cover a large fraction of the solar disk around aphelion (see Figure~\ref{fig:FOV}).

The observing strategies for the science goals are summarized in Table~\ref{tab:science}. More details for the individual science objectives are given in the following subsections. For all science objectives, maps of the magnetic field should be transferred from time to time to provide context information. We expect that a duty cycle greater than 80\% may be acceptable for helioseismology.

Some of the science goals are not within the limitations of the baseline telemetry allocation and observing windows (three science windows per orbit). As discussed in Section~\ref{sec.compression}, it might be possible to decrease the required telemetry by compressing the data. However, reducing the observing duration would severely affect the signal-to-noise ratio. Probing large-scale flows is affected by the realization noise of the solar oscillations and noise arising from supergranulation. Both decrease with $T^{-1/2}$, where $T$ is the observing time.

In addition to the observing duration and telemetry, many of the science goals described below have some requirements regarding the orbit. This is especially the case for the measurements at high latitudes. These should be performed while the spacecraft is at high latitudes. During the extended mission, {\it Solar Orbiter} will reach a maximum solar latitude of $34^\circ$ and will be above $30^\circ$ for up to 23 consecutive days. This is comparable to the required observing time for high-latitude measurements.

We note that helioseismology of PHI data will require some (potentially complicated) modifications to standard methods to account for the time-varying image geometry.

\begin{sidewaystable}
\caption{Table of helioseismology science objectives. Some science goals exceed the currently allocated observing time or telemetry (highlighted in red, green colors indicate that the parameter is within the allocation)}
\label{tab:science}
\begin{tabular}{p{7cm}p{2.4cm}p{2.3cm}p{3.3cm}p{2.5cm}}
\hline\noalign{\smallskip}
Science target & \# Spatial points & Observing time & Observables & Approx. telemetry \newline (5 bits/observable)\\
\noalign{\smallskip}\hline\noalign{\smallskip}
{\bf Near-surface rotation, meridional circulation, and solar-cycle variations at high latitudes} &  &  & &\\
- Helioseismology & $512\times 512$ & \cellcolor{red} 30 days & $\vlos$ every 60 s & \cellcolor{red} 60 Gbit\\
- Solar-cycle variations from helioseismology & $512\times 512$ & \cellcolor{red} $4\times 30$~days, 2 years apart & $\vlos$ every 60 s& \cellcolor{red} $4\times 60$~Gbit\\
- Meridional circulation to 3 m/s at $75^\circ$ (see Fig.~\ref{fig:sg_noise})& $512\times 512$ & \cellcolor{red} 100+ days & $\vlos$, $I_{\rm c}$ \& $B_{\rm los}$ & method dependent\\
- Granulation and magnetic-feature tracking & $2048\times 2048$ & \cellcolor{red} 30 days & $I_{\rm c}$ \& $B_{\rm los}$, two consecutive images every 8 h & \cellcolor{green} 8 Gbit\\
- Supergranulation tracking & $512\times 512$ & \cellcolor{red} 30 days & $\vlos$ every 60 min & \cellcolor{green} 1 Gbit\\
\hline
{\bf Deep and large-scale solar dynamics } & & & & \\
- MDI-like medium-$l$ program & $128\times 128$ & \cellcolor{red} continuous & $\vlos$ every 60 s& \cellcolor{green} 40 Gbit/year \\
- Stereoscopic helioseismology (PHI + other instrument) & $128\times 128$ & \cellcolor{red} continuous & $\vlos$ every 60 s& \cellcolor{green} 40 Gbit/year \\
\hline
{\bf Convection at high latitudes} & && &\\
- Helioseismology & $1024\times 1024$ & \cellcolor{green} 7 days & $\vlos$ every 60 s & \cellcolor{green} 50 Gbit\\
- Feature tracking & $2048\times 2048$ & \cellcolor{green} 7 days & $I_{\rm c}$ \& $B_{\rm los}$, two consecutive images every 8 h & \cellcolor{green} 2 Gbit\\
\hline
{\bf Deep convection and giant cells} & & & & \\
- Helioseismology & $128\times 128$ & \cellcolor{red} continuous & $\vlos$ every 60 s& \cellcolor{green} 40 Gbit/year \\
- Feature tracking & $512\times 512$ & \cellcolor{red} $4\times 60$~days & $\vlos$ every 60 min & \cellcolor{green} $4\times 2$ Gbit\\
\hline
{\bf Active regions and sunspots} & & & & \\
- Active region flows \& structure & $512\times 512$ & \cellcolor{red} 20 days &  $\vlos$ every 60 s & \cellcolor{green} 40 Gbit\\
- Sunspot oscillations & $1024\times 1024$ & \cellcolor{green} 2 days & $\vlos$, $I_{\rm c}$ \& $\vec{B}$ every 60 s & \cellcolor{red} 80 Gbit\\
- Calibration far-side helioseismology & $128\times 128$ & \cellcolor{green} $5\times 2$~days & $B_{\rm los}$, $I_{\rm c}$, \& $\vlos$ every 60 s& \cellcolor{green} $5\times 0.3$~Gbit\\
\hline
{\bf Physics of oscillations (stereoscopic obs.)} & & & &\\
- Effect of granulation on oscillations & $2048\times 256$ & \cellcolor{green} 1 day & 6 filtergrams every 60 s & \cellcolor{green} 20 Gbit\\
- Two components of velocity & $512\times 512$ & \cellcolor{green} 10 days & $\vlos$ every 60 s & \cellcolor{green} 20 Gbit\\
- Magnetic oscillations & $2048\times 2048$ & \cellcolor{green} 1 day & $\vlos$, $I_{\rm c}$, $B_{\rm los}$ every 60 s \& $\vec{B}$ at max. cadence & \cellcolor{red} 100 Gbit\\
\hline
{\bf Low resolution observations} & & & &\\
- LOI-like observations (solar-cycle variations, active longitudes) & $4\times 4$ for $\vlos$ \& $I_{\rm c}$, $32\times 32$ for $B_{\rm los}$& \cellcolor{red} continuous& $\vlos$ \& $I_{\rm c}$ every 60 s \& $B_{\rm los}$ once per day & \cellcolor{green} $0.1$~Gbit/year\\
- Shape of the Sun & $10\times 6000$ & \cellcolor{green} Every few months & $I_{\rm c}$ at 12 angles during rolls & \cellcolor{green} 4 Mbit for one roll\\
\noalign{\smallskip}\hline
\end{tabular}
\end{sidewaystable}

\begin{figure}[h!]
\floatbox[{\capbeside\thisfloatsetup{capbesideposition={right,top},capbesidewidth=6cm}}]{figure}[\FBwidth]
{\caption{Noise level due to supergranulation in a surface average of rotation or meridional circulation over longitude in 30 days ({\it solid curve}) and 120 days ({\it dashed curve}) of continuous measurements as a function of heliographic latitude, $\lambda$.   The noise level for an observing time of 30~days scales as $\sim (1 \; {\rm m/s}) \times (\cos{\lambda})^{-1/2}$, where 1~m/s is the typical noise level at the equator~\citep[Figure 6  of ][]{2001ESASP.493..227G}, and $\cos{\lambda}$ is the ratio of the supergranule number between the equator and latitude $\lambda$.  The same estimate follows from simple assumptions for rms amplitude, lifetime, and length scale of the supergranulation pattern.  In order to reduce the noise from supergranulation, time averaging (or time sampling) is required.  The noise due to solar convective flows is not only important for helioseismology, but must also be considered for correlation tracking and other surface measurements} \label{fig:sg_noise}}
{\includegraphics[width=5cm]{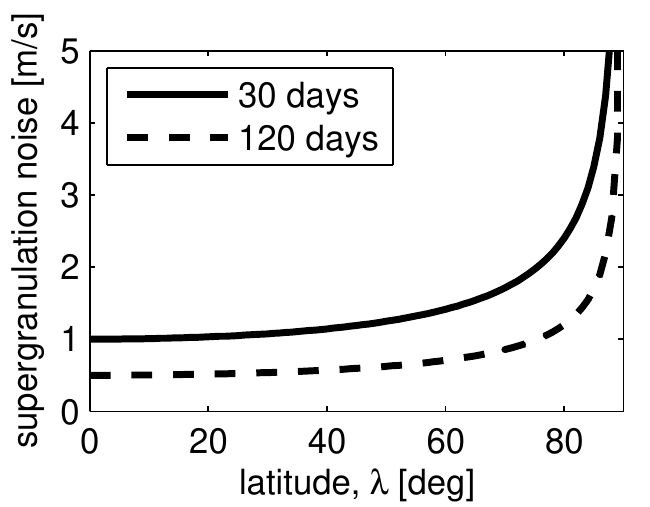}}
\end{figure}


\subsection{Near-Surface Rotation, Meridional Circulation, and Solar-Cycle Variations at High Latitudes}

One major science objective of {\it Solar Orbiter} is to understand large scale flows near the solar poles, which are in turn important for understanding solar dynamics and the solar cycle.  Differential rotation and meridional circulation are essential components in dynamo models, especially flux transport dynamos.    The importance of these large scale flows at high latitudes has been discussed by, e.g., \citet{2009ApJ...693L..96J,2012ApJ...746...65D}.  While helioseismology has mapped differential rotation in most of the solar interior (e.g., see the review by~\cite{2009LRSP....6....1H} and references therein), the cycle dependence of differential rotation (``torsional oscillations'', $> 20$~m/s) is not well measured for the full Hale cycle at latitudes above about sixty degrees.  Variations from one sunspot cycle to the next have been seen in the differential rotation at high latitude~\citep{2009SPD....40.2401H}.
The Sun might also exhibit a polar vortex~\citep{1979ApJ...231..284G}, as already seen on planets~\citep[e.g., ][]{2008Sci...319...79F}. In some dynamo models, the meridional circulation sets the cycle period and thus is an important target.  The meridional circulation has an amplitude of less than fifteen m/s at mid-latitudes (and less at high latitudes) and is thus a challenging objective.  Progress has been made using local helioseismology~\citep{2013ApJ...774L..29Z}, although a lot of uncertainty remains in the deep convection zone and at high latitudes.


Near the surface, the limiting factor in the measurement of longitudinally-averaged flows is the noise introduced by the supergranulation flows ($\sim200$~m/s rms).  The noise introduced by supergranulation flows goes up with latitude, as the number of supergranules at fixed latitude goes down.  This is illustrated in Figure~\ref{fig:sg_noise}, which is based on a very simple model that does not take account of the inclination of the orbit.  Time averaging over long periods is necessary to measure longitudinally-averaged flows at high latitudes~\citep[e.g.][]{2013ApJ...767L..20H,2013SoPh..287...85K}.  One month of data will result in a noise level of two m/s at a latitude of $75^\circ$, using time-distance helioseismology.  A noise level of one m/s at $75^\circ$ would require $\sim 120$ days of observations. These noise estimates are not specific to the particular method employed (helioseismology or other); it is inherently due to noise from solar convective flows.  One m/s at $85^\circ$ would require 
continuous 
coverage for a significant fraction of the mission. Standard local helioseismology requires spatially resolved observations with resolution of at least $512\times 512$ pixels across the solar disk.  All helioseismology applications require a cadence of at least 60~s.  In order to look for solar-cycle variations of these large-scale flows, we suggest to repeat these observations several times during the mission. 

It should be noted that the helioseismology observations (e.g., $\vlos$ or $I_{\rm c}$) will be used for non-helioseismic purposes.
For example, flows at the surface can also be determined by correlation tracking of granulation or supergranulation.
Granulation tracking requires observations at full spatial resolution  \citep[e.g.][]{2013A&A...552A.113R} and pairs of images taken 60~s apart, with at least a few pairs taken per day. 

Another way to determine the flows near the surface is to track the supergranulation in Dopplergrams~\citep[e.g.][]{Hathaway2013}. This has the advantage that only temporally averaged Dopplergrams with low cadence (one hour) and modest resolution ($512\times 512$ is known to be sufficient, but $256\times 256$ may be acceptable) are needed. \citet{2003ApJ...596L.259S} used 60 days of data to detect the meridional flow up to about $\sim 75^\circ$ latitude, but without resolution in depth. 

\subsection{Deep and Large-Scale Solar Dynamics}

One of the main science goals of {\it Solar Orbiter} is to measure flows in the deep interior of the Sun. These are currently not well characterized. Recently, \citet{2013ApJ...774L..29Z} reported the discovery of multiple cells of the meridional circulation throughout the convection zone but high latitudes have not been explored. In addition, the $1.3$-year oscillations in the rotation rate near the tachocline~\citep{Howe2000} remain a puzzling result.  These oscillations have an amplitude that increases with latitude and could be measured with {\it Solar Orbiter}.

Only low spatial resolution is needed for deeper flows.  On the other hand, the measurement of flows in the deep interior requires long time-series.  \citet{2008ApJ...689L.161B} estimated that up to ten years would be required to measure a few m/s return meridional flow at the base of the convection zone. However, a noise level of 5 m/s could already be achieved with 6 months of data.


By combining observations from {\it Solar Orbiter} with those from space or the ground to maximize the spatial coverage, improvements are expected.  Thanks to its unique orbit, {\it Solar Orbiter} will be the first mission to study the advantages of stereoscopic helioseismology. For the purpose of stereoscopic helioseismology of the deep interior, observations with a resolution of $128\times 128$ should be sufficient. Observing the Sun from opposing angles offers the opportunity to perform time-distance helioseismology using ray-paths probing the deep interior of the Sun. This method has never been used before and may lead to new discoveries.  Other than the preliminary estimates of~\citet{2003ESASP.517...71R},  models have not yet been developed to predict the signal-to-noise ratios that will be possible using this method. For this science target, the duty cycle should be above 80\%. The data for this represents only a fraction of the 
currently allocated telemetry, but does require observations over very long periods of time.


 

\subsection{Convection at High Latitudes}

Another major objective is to understand the properties of convection, in particular supergranulation, which almost 60 years after its discovery still remains enigmatic (e.g. preferred length scale of convection, wavelike properties). This in turn will allow us to better understand issues such as flux transport and thus the solar dynamo.

Supergranulation is one of the strongest signals in time-distance helioseismology and will no doubt be a feasible science target for PHI within the baseline resources. There are several reasons to expect an unusual behavior of the supergranulation at high latitudes. According to \citet{Nagashima2011} supergranules may preferentially align in the North-South direction at high latitudes.  Furthermore, the effects of the Coriolis force on supergranulation flows \citep[see][]{2010ARA&A..48..289G} is expected to be strongest in the polar regions. Finally, the wavelike properties reported for supergranulation~\citep{2003Natur.421...43G,2003ApJ...596L.259S} show a clear latitude dependence.

A snapshot of supergranulation flows near the surface can be obtained using local helioseismology with a few hours of observations of moderate spatial resolution ($1024\times 1024$ pixels). The study of the evolution of the supergranulation pattern requires at least a week of data (a few lifetimes). In addition to helioseismology, local correlation tracking with high resolution ($2048\times 2048$) but low cadence (two consecutive images every 8 hours) could be used to study the supergranulation flows at the surface.

\subsection{Deep Convection and Giant Cells}

Recent helioseismology results call into question current models of deep solar convection.  \citet{2012PNAS..10911928H} found much lower flow velocities for deep convection than predicted.  Extending this work with {\it Solar Orbiter} would require continuous observations over a large fraction of the mission, with a spatial resolution of $128\times 128$~pixels.

Recently,~\citet{Hathaway2013}, using correlation tracking, reported the detection of persistent giant cells at the solar surface, but were unable to observe close to the poles. Giant cells may play an important role in transporting angular momentum. This presents a new and exciting opportunity for PHI. For studying giant cells, we will need several months of observations consisting of $512\times 512$~pixels with a cadence of 60~min~\citep{Hathaway2013}. This could either be achieved using a single but very long time-series (of order a full orbit) or using a few moderately long (2-3 months) time-series with a similar combined length. It may also be desirable to repeat this after a substantial period to look for any solar-cycle dependence in the giant cell pattern.   This analysis could also be done with helioseismology over the same observing time, but with a cadence of 60~s.

\subsection{Active Regions and Sunspots}

Another interesting possibility is to be able to increase the temporal coverage of evolving active regions using both helioseismology and correlation tracking, as well as to observe the flux dispersal over longer times than currently possible. This will require 20 days of data (20 days in combination with, e.g., HMI observations would allow continuous observations of the active region for more than a full solar rotation) with a cadence of 60 s and a resolution of $512\times 512$ pixels.

Stereoscopic observations would also greatly benefit sunspot seismology since it would be possible to observe two components of MHD waves. These observations will likely not require extensive observations. Previous studies~\citep[e.g.][]{Schunker2005,2006ApJ...643.1317Z} were able to probe sunspots with local helioseismology using MDI data with a length of a few days. We estimate that 2 days of data should be sufficient with a spatial resolution of $1024\times 1024$ pixels. However, in order to study MHD waves, in addition to the Doppler velocity, the vector magnetic field and the continuum intensity will be required. Coordination with an additional instrument will be necessary, meaning that the active regions being investigated will need to be on the nearside of the Sun.

Since {\it Solar Orbiter} will be able to observe the farside of the Sun, it can be used to calibrate farside helioseismology~\citep{2000Sci...287.1799L}. Active regions not visible from the Earth may be imaged with {\it Solar Orbiter} ($B_{\rm los}$, $I_{\rm c}$) and compared with helioseismic farside maps derived from a near-Earth vantage point. In addition, combining {\it Solar Orbiter} and other helioseismology datasets would improve farside imaging (e.g., using different skip geometries). Since farside helioseismology relies on medium-degree modes, this would be possible with a few days of data obtained for the low resolution helioseismology ($128\times 128$ pixels).


\subsection{Physics of Oscillations (Stereoscopic Observations)}
Vector velocities, utilizing two vantage points to observe the same location, also present exciting possibilities. Recent attempts at measuring deep meridional flows using local helioseismology are affected by apparent center-to-limb effects~\citep{2012ApJ...749L...5Z}, which are suspected to be caused by asymmetries in granular flows~\citep{2012ApJ...760L...1B}. Stereoscopic observations would allow the evaluation of the dependence of this effect on the observation angle. We suggest to observe a strip with a size of $2048\times 256$ pixels, covering all heliocentric angles, for about a day. In order to study the dependence of this effect with height in the atmosphere, the filtergrams could be transmitted instead of Dopplergrams obtained onboard.

Obtaining vector velocities would allow observations of the relation between radial and horizontal velocities in supergranulation or the phase relationships between different velocity components of helioseismic waves near the solar surface and thus give insight to the physics of p- and f-modes. Observing supergranulation does not require a high spatial resolution (e.g. $512\times 512$ pixels) but it is necessary to observe for several lifetimes of the supergranulation (e.g. 10 days).

In addition, vector velocities in combination with the full magnetic field vector would give insight to MHD waves in active regions~\citep[e.g.][]{2000SoPh..192..403N}. This would require stereoscopic observations of an active region with high spatial resolution ($2048\times 2048$ pixels). \citet{2000SoPh..192..403N} used a few hours of data, so we estimate that an observing length of the order of one day should be sufficient here.


\subsection{Low Resolution Observations}
Additional helioseismology observation modes that should be considered are unresolved observations or observations with very low resolution. 
Sun-as-a-star time-series from \newline SOHO/VIRGO~\citep{1995SoPh..162..101F} are used regularly for testing concepts for asteroseismic data analysis~\citep[e.g.,][]{2009AIPC.1170..560G, 2010arXiv1003.5178S,2013PNAS..11013267G}. PHI would provide a unique data set in this regard as it would allow observations from outside of the ecliptic plane. For this purpose (nearly) continuous observations ($I_{\rm c}$ and $\vlos$)  with very low resolution (e.g. $4\times 4$ pixels) using the FDT would be appropriate.

Another case for observations using only few pixels  is the study of the shape of the Sun~\citep[e.g.,][]{2012Sci...337.1638K}.  This would require rolls of the spacecraft, which perhaps would have to be coordinated with calibrations of the in-situ instruments on {\it Solar Orbiter}. In order to save telemetry, only a band of pixels along the limb would have to be transmitted.

\section{Helioseismology: Selected Requirements}

\subsection{Example Telemetry Requirements}\label{sec.example_telem_calcs}

The last column of Table~\ref{tab:science} shows a set of example telemetry estimates made using some simple assumptions. Here we have used 5 bits/pixel, regardless of the type of observable or cadence and have not distinguished between the two telescopes (FDT or HRT). For some objectives it is possible to only transmit parts of the images (e.g. the parts that are not seen from the Earth), but this has not been accounted for. In addition, we have not taken into account the number of pixels that are off the solar disk when computing the data rates for full-disk images. Given that the spacecraft-Sun distance will be changing it will be necessary to have variable binning and extraction as a function of time, but here we have simply assumed that the images have the desired size.

Some science objectives in the Table are within the baseline of telemetry~\citep[51 Gbit/orbit,][]{Redbook} and others require more telemetry. We are currently investigating ways (compression algorithms) to optimize data rates. As will be discussed in Sect.~\ref{sec.compression} there are many possibilities for reducing the amount of telemetry and a factor of two to five gain in compression efficiency may be possible. These estimates are based on the experience on compression from previous space missions, as for example MDI~\citep{1995SoPh..162..129S,1997SoPh..170...43K}.

One promising avenue for refining the very simplistic example estimates shown here is direct numerical simulation of both wave propagation and synthetic PHI observations (see Sect.~\ref{sect:SOPHISM} and~\ref{sect:synth_data}). In particular for the case of stereoscopic helioseismology of the tachocline, it will be useful to have wave propagation simulations in spherical geometry. These simulations will help in the exploration of methods for getting the most out of stereoscopic or other helioseismic observations.

An important consideration in planning observations is that many of the most important phenomena on the Sun change with time. Snapshots of solar interior structure and dynamics are important but are not as informative as continuous coverage. In the ideal case, we would be able to track the temporal evolution of flows at high latitudes and in the tachocline throughout the {\it Solar Orbiter} mission. For this reason, it may be important to consider the value of carrying out low data rate but continuous observations in addition to, or in place of, a few short intervals of high data-rate observations. This would, of course, require observations outside of the science windows.

\subsection{Thoughts on Compression}\label{sec.compression}
The issue of how to best trade off data coverage versus compression is a complex one. Obviously good temporal coverage at high cadence, with good S/N and good resolution is desired, but the telemetry is limited and so something has to give. In some cases, such as reducing the length of observations, it is fairly straightforward to evaluate the impacts. For many other options, it will probably be necessary to simulate data or use existing data from other instruments, which can then be degraded. This study is underway~\citep{compression}.

\subsubsection{Binning, Subsampling, and Cropping}

These methods are perhaps the simplest ways to reduce the amount of data. Indeed, these were the main methods applied for MDI. Straight binning and subsampling tends to be neither efficient from a compression point of view, nor optimal from a science point of view. A large amount of spatial aliasing is generated leading to artifacts on e.g.\ power spectra and the data are not efficiently compressed. To overcome these problems, the data should be smoothed with a low-pass filter before subsampling. An example of how to do this is the MDI Medium-$l$ program where the images are convolved with a 2D Gaussian before subsampling, which greatly reduces the amount of aliasing and associated  artifacts~\citep{1996kosovichev,1997SoPh..170...43K}. Note that filtering the data leads to a significant reduction in the compressed size, even in the absence of subsampling, at least for most lossless compression algorithms.

The MDI Medium-$l$ program also crops the images at $0.9 \ R_\odot$. Even more substantial reductions are also possible, such as was done in some MDI campaigns where only a subsection of the images were downlinked. While the MDI Medium-$l$ program has been extremely successful, it is also clear that a better trade-off could have been made. The images were probably too severely cropped and the anti-aliasing filter could have been better designed. Originally it was planned to do an onboard Spherical Harmonic Transform (SHT) instead of the Medium-$l$ program, but this was not done for a variety of reasons. The advantage of such a scheme (onboard SHT or similar) is that it allows for filtering that is more uniform in physical scale on the Sun than does a simple smoothing and subsampling. While impossible given the MDI hardware, it would also be possible to apply a spatially variable filtering, which would have a similar effect. For PHI it may be worth it to consider some of these options if the hardware and 
software allow.

\subsubsection{Truncation}
This is another simple way of reducing the amount of data. The data are divided by a scale factor and then rounded. As long as the scale factor is small compared to the range of the data, this is almost equivalent to adding white noise.  Such a rounding comparable to the photon noise is certainly reasonable, but the effect of larger numbers is currently being evaluated.

In the case of intensity data it may make sense to pass the data through a square root lookup table instead. This results in a constant fractional increase in the noise relative to the photon noise. This scheme is used for HMI and MDI.

\subsubsection{Lossless Compression}

For both MDI and HMI the compression (beyond truncation and sqrt lookup) was lossless and in both cases done by first differencing the numbers in the readout direction (roughly parallel to the equator) and passing the result through a Rice compression. This was, of course, combined with transmitting the first pixel and various data to prevent excessive data loss in case one packet is lost.

A compression scheme like this can be considered to consist of two parts: a prediction of the current pixel (it is the same as the previous one) and an encoding of the difference. This view leads to two obvious directions for improvement: better prediction and better compression of the differences. For the prediction several improvements can be made. Instead of using the preceding value one can use multiple preceding values. Since the data have significant high frequency power, this does not lead to a very large improvement. A further improvement can be made by looking in two dimensions, e.g.\ by using the average of the nearest preceding pixels in x and y (both of which have already been transmitted). This leads to a modest improvement and is similar to what is done in the standard lossless JPEG algorithm~\citep{Wallace1992}. This can also be extended to using the preceding image(s) in time, which in some case can lead to a significant improvement. It may be noted that using a substantial number of images 
back in time is more or less equivalent to numerically propagating the observed wavefield.

Rice compression has similar performance to a Huffman coding~\citep{Huffman}, which is in turn optimal if the probabilities are known and we restrict ourselves to encoding one difference at a time. Obvious improvements on a Huffman coding include making it depend on the position in the image, using longer symbol length and going to an arithmetic coding.

Finally it may be possible to use standard lossless compression algorithms, like various LZ algorithms~\citep{Ziv1977,Ziv1978}, JPEG and so forth.

\subsubsection{Lossy Compression}

It may also be worth it to test various lossy algorithms, like JPEG. However, there are some serious issues to consider regarding this. First of all the oscillations look essentially like white noise (filtered by the PSF) in a single image. This makes it very difficult to compress away noise while keeping the signal. Another problem is that algorithms such as lossy JPEG divide images into (e.g. $8\times 8$ pixel) blocks. Likely this will cause strong artifacts at the scale of the blocks.

\subsubsection{Some Further Thoughts}

It may also be possible to employ further tricks to reduce the data volume. One approach is to use more advanced filtering than a simple convolution by a smooth (e.g. Gaussian) kernel. SHTs and variable width Gaussians are some possibilities, but there may be other options. More interesting is to also use the time dimension. One simple possibility is to suppress the signal at low (granulation) and high (above acoustic cutoff) temporal frequencies. But in principle more advanced filtering schemes could be employed on full 3D transforms. Similarly it may be possible to encode the transform directly instead of going back into the space-time domain. Transmitting SHT or temporal Fourier Transform coefficients is one possibility, but others exist.

Clearly some of the various ideas are easy to test. In particular the lossless compression algorithms where the success criteria are straightforward (i.e. the compression efficiency and computational cost). Others are more tricky as they involve evaluating if the damage introduced by various lossy algorithms is acceptable. We are currently working on evaluating some of these methods.

\subsection{Other Requirements}

In addition to the requirements on duration, resolution, spatial and temporal coverage already mentioned, several other things are needed to make optimal use of the data.

First of all it is important that the image geometry is well known. The image scale and pointing for the FDT telescope should be derived onboard using the solar limb. For the HRT this is more difficult, as the limb is not in general visible. For that it will be necessary to cross-correlate FDT and HRT images. Unfortunately the offset and scale difference is unlikely to be constant. There will be long term drifts due to thermal effects and so measurements will have to be made under various conditions. Also, the offset has a certain random component which will have to be characterized.

Some of the science objectives are achieved with FDT data for which the requirements are easily met. A subset of the rest of the objectives either involve studies where the absolute pointing is not critical (e.g. convection in polar regions, active region science) or ones where we have a simultaneous view from HMI (e.g. observations of the same area from different directions). For some science objectives, it may be necessary to observe the limb in the HRT field of view.

A more difficult requirement is that of knowing the roll angle of the instrument. To avoid that the solar rotation leaks into the measurement of the meridional flow, the absolute roll must be known to about one arcmin. Likely the only way this can be achieved is by cross-correlating with HMI or a ground-based instrument, but this will have to be repeated to check for any thermally induced variations. The roll cannot be determined using the star-trackers of the spacecraft because there might be an offset between the instrument and the star-tracker which needs to be determined. This offset could depend on temperature and, thus, change during the orbit. HMI and Solar Orbiter do not need to have the same viewing angle for this purpose. It will be sufficient if both spacecraft see the same region on the Sun.

Some further thoughts based on lessons learned from MDI and HMI are in section~\ref{sect:lessons}.

\section{Polarimetric and Helioseismic Imager}
\label{sect:PHI}

\subsection{Observables}

PHI will obtain two-dimensional intensity images for six wavelength points within the Fe I $6173$~\AA \ line, while measuring four polarization states at each wavelength. The current baseline is to measure at five wavelength positions within the line ($-160$, $-80$, $0$, $80$, and $160$~m\AA \ from the line center) and at one continuum point ($-400$~m\AA \ from the line center).

\subsection{Instrument Concept and Implementation}
\begin{figure}[h!]
 \includegraphics[width=\textwidth]{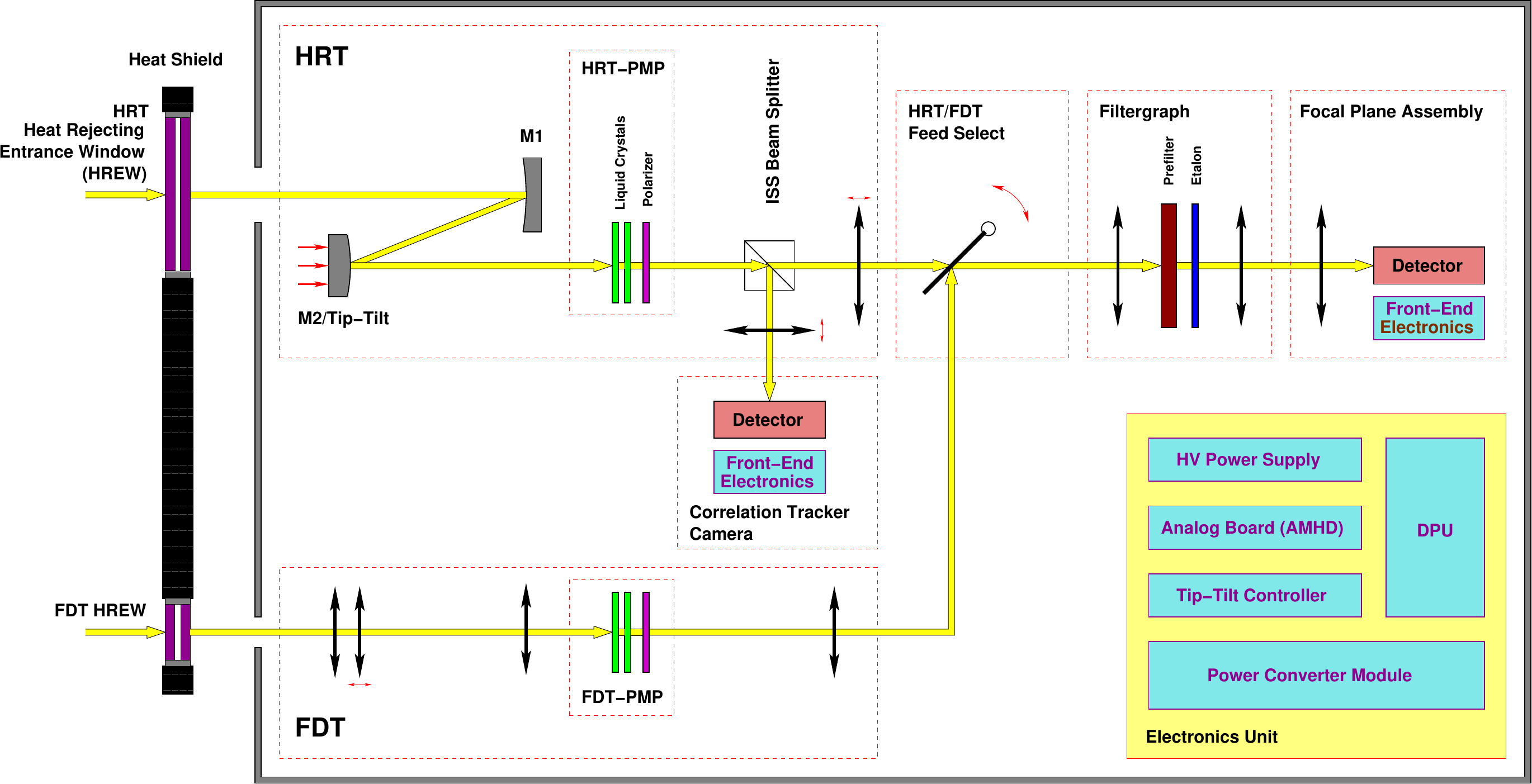}
\caption{Schematic drawing of the PHI instrument. PHI will consist of two telescopes, the High Resolution Telescope (HRT, {\it upper left part of the diagram}) and the Full Disk Telescope (FDT, {\it lower left part of the diagram}). The HRT will have its own Image Stabilization System (ISS) consisting of a tip-tilt mirror and a correlation tracker. Both telescopes have individual Polarization Modulation Packages (PMPs) but feed the same filtergraph and one detector. For more details about the instrument see text}
\label{fig:diagram}
\end{figure}

PHI will consist of two telescopes, which feed one filtergraph and one focal plane array: the High Resolution Telescope (HRT) is designed as an off-axis Ritchey-Chr\'{e}tien telescope with a decentered pupil of 140~mm diameter. The HRT includes an internal image stabilization system (ISS) based on a fast steerable mirror that reduces residual pointing errors by the spacecraft to levels compatible with high resolution polarimetry. The error signal is derived from a correlation tracker camera inside the HRT. The Full Disk Telescope (FDT) is designed as a refractive telescope with $17$~mm aperture diameter. The two telescopes will be used sequentially. A schematic drawing of PHI is given in Figure~\ref{fig:diagram}.

Both telescope apertures are protected from intense solar flux by special Heat Rejecting Entrance Windows (HREW), which are part of the heat-shield assembly of the spacecraft. They are purely dielectric broad-band reflectors with a narrow notch in the reflectivity curve around the science wavelength of the instruments. With more than 80\% transmittance at the science wavelength, in combination with almost perfect blocking from 200~nm to the far infrared, the heat load into the instrument can be effectively decreased, while preserving the high photometric and polarimetric accuracy of PHI.

The polarimetric analysis is performed by two Polarization Modulation Packages (PMPs), one in each of the telescopes. Each PMP consists of two nematic liquid crystal variable retarders (LCVRs), followed by a linear polarizer as analyzer. The modulation scheme is the same as the one used in the Imaging Magnetograph eXperiment~\citep[IMaX,][]{2011SoPh..268...57M} onboard the successful Sunrise balloon-borne observatory~\citep{2010ApJ...723L.127S,2011SoPh..268....1B}.

\begin{figure}
\begin{minipage}[h!]{\textwidth}
\includegraphics[width=0.5\textwidth]{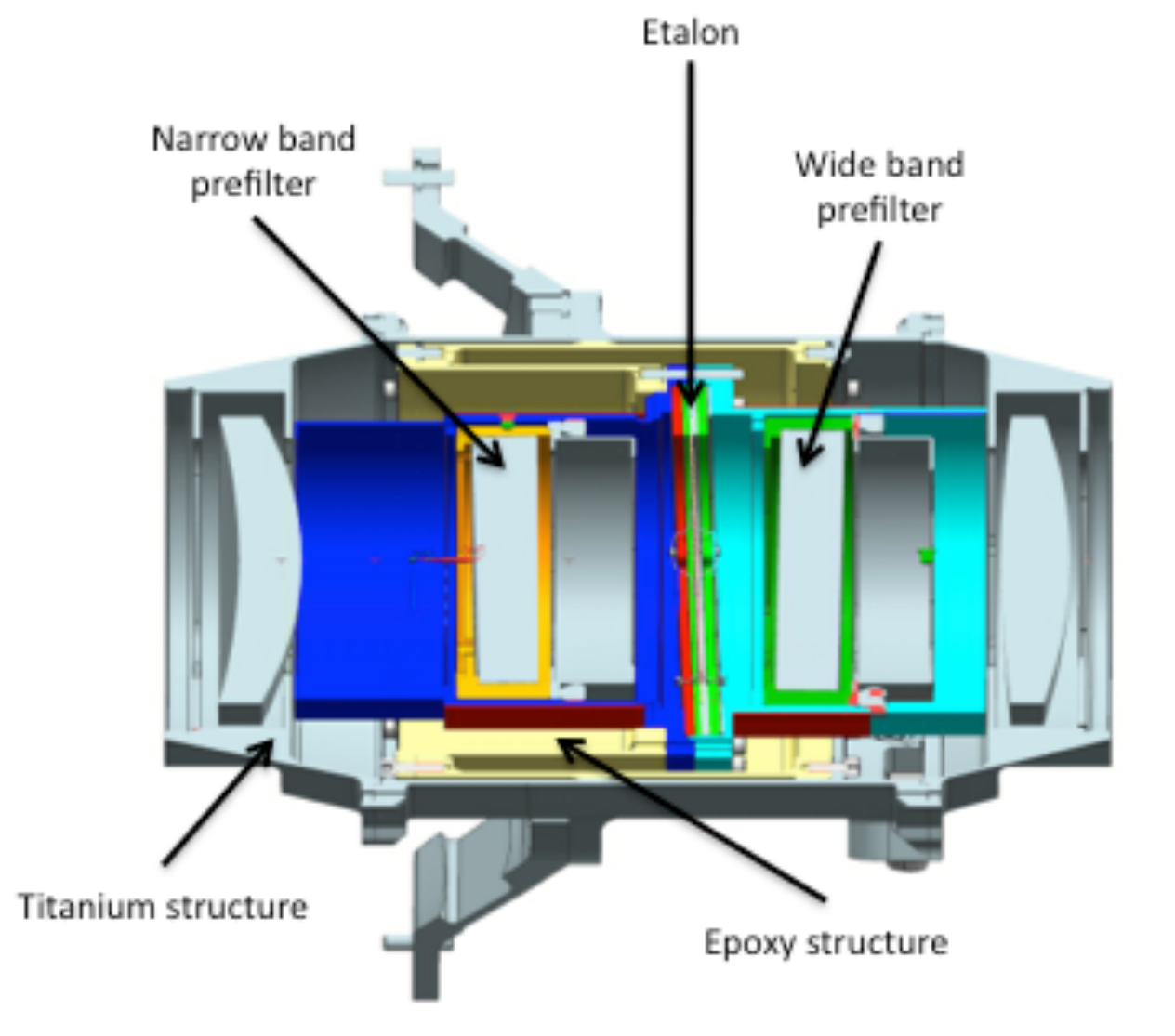}
\includegraphics[width=0.5\textwidth]{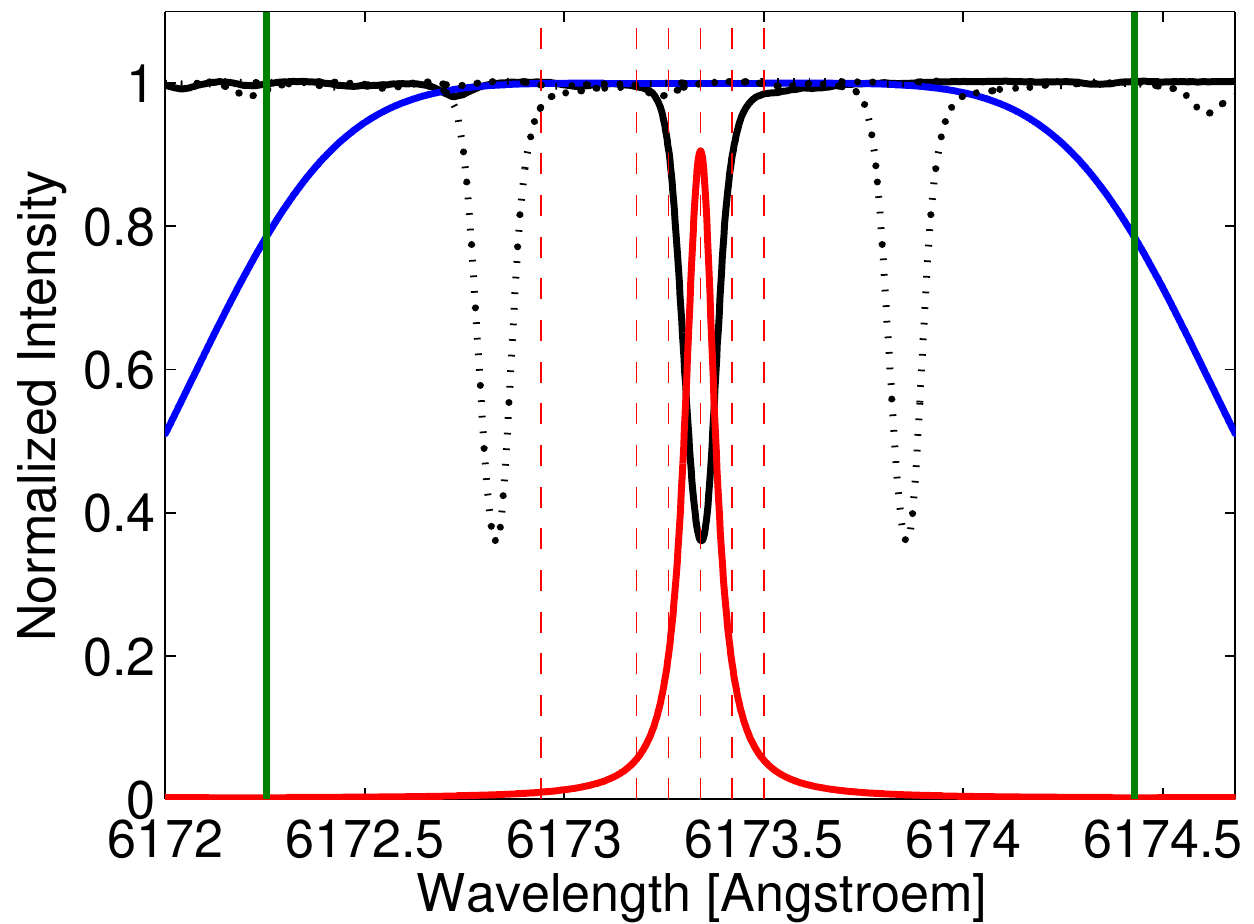}
\end{minipage}
\caption{{\it Left}: Cut of the filtergraph. The filtergraph consists of a freely tunable etalon in a telecentric mounting and two prefilters on either side of the etalon, a narrow-band prefilter and a wide-band prefilter. {\it Right}: Schematic representation of the filter curves of PHI. The Fe I $6173$~\AA \ line ({\it solid black line}) will be scanned by using three filters, the wide-band prefilter (not shown here), the order-selecting narrow-band prefilter ({\it blue curve}) and a freely tunable etalon ({\it example profile shown by the red curve}). The current baseline is to scan the line at six wavelength positions ({\it vertical dotted lines}). Due to the large orbital motion of the spacecraft (up to 25 km/s), the line will be subject to a large Dopplershift (up to $\pm 0.5$~\AA, {\it denoted by the dotted black lines}). The etalon has a sufficiently large tuning range to account for this ({\it given by the solid green lines})}
\label{fig:filtergraph}
\end{figure}

The filtergraph (see Figure~\ref{fig:filtergraph}) for spectral analysis consists of a lithium niobate (LiNbO$_3$) solid state etalon that can scan the line over about 2 \AA \ (the line is 120 m\AA \ wide). Its optical design is based on heritage from the IMaX instrument, while the filtergraph structure is specifically designed for {\it Solar Orbiter} by IAS in Orsay. The transmission curve of the etalon exhibits evenly distributed peaks separated by the product of the index of refraction of the lithium niobate and the thickness of the etalon (typically about 250 $\mu$m). In order to select the correct transmission peak, the etalon is used in tandem with a set of two prefilters: the wide-band prefilter and the narrow-band prefilter having a Full Width at Half Maximum (FWHM) of 100 \AA \ and 2.7 \AA \, respectively (see right part of Figure~\ref{fig:filtergraph}). Before the light enters the etalon, the narrow-band prefilter selects the working-order of the etalon. This filter also transmits green light that 
is used for ground-based calibrations. The green light is removed by the narrow-band prefilter, located behind the etalon. The tuning over the line is performed by using the piezoelectric effect of the lithium niobate, affecting both the thickness and the index of refraction. A change of $\sim 6$~kV allows tuning the etalon over 2 \AA. The width of the narrow-band prefilter and the tuning range of the etalon are sufficient to account for Dopplershifts of the line caused by the orbit (up to $\pm 0.5$~\AA).


The etalon is used in a telecentric beam, i.e. all the points in the solar image see a cone of light with the same opening impinging on the etalon at normal incidence angle. The stability of the filtergraph is driven by two main design constraints: the thermal stability for periods shorter than 30 min, and by the wavelength repeatability induced by the high voltage power supply. The thermal design of the filtergraph ensures that the inertia is about 7 hours, thereby making the instrument 
extremely stable to variations from the environment. A stability better than $0.01^\circ$~C peak-to-peak for periods shorter than 30 min was measured during thermal tests of the breadboard corresponding to 20 m/s; this should allow the measurement of modes with degree higher than four. The high voltage repeatability was also measured in the lab to value of 3 m/s, this is 50 times lower than specified.


The focal plane assembly is built around a $2048\times 2048$ pixel Active Pixel Sensor (APS), which is especially designed and manufactured for the instrument. It will deliver 10 frames per second which are read out in synchronism with the switching of the polarization modulators. In order to increase the signal-to-noise ratio, PHI will accumulate several images for every wavelength position and polarization state.


\section{Simulation Tool: SOPHISM}
\label{sect:SOPHISM}

SOPHISM is a software simulator aimed at a full representation of the PHI instrument (both hardware and software) and is applicable to both telescopes of the instrument. Starting from 2D maps of line profiles for the Fe I $6173$~\AA\ line computed from MHD simulations, SOPHISM generates synthetic 2D maps of the observables that will be measured by PHI. This allows estimates of the performance of PHI, which helps optimize the instrument design and the onboard processing. The hardware part of the simulations takes into account all the elements affecting solar light from when it enters the telescope (including some perturbations due to the spacecraft) up to the registering of this light on the detector. The software part deals with the data pipeline and processing that takes place onboard.

\begin{figure}
\begin{minipage}[h!]{\textwidth}
\includegraphics[width=0.5\textwidth]{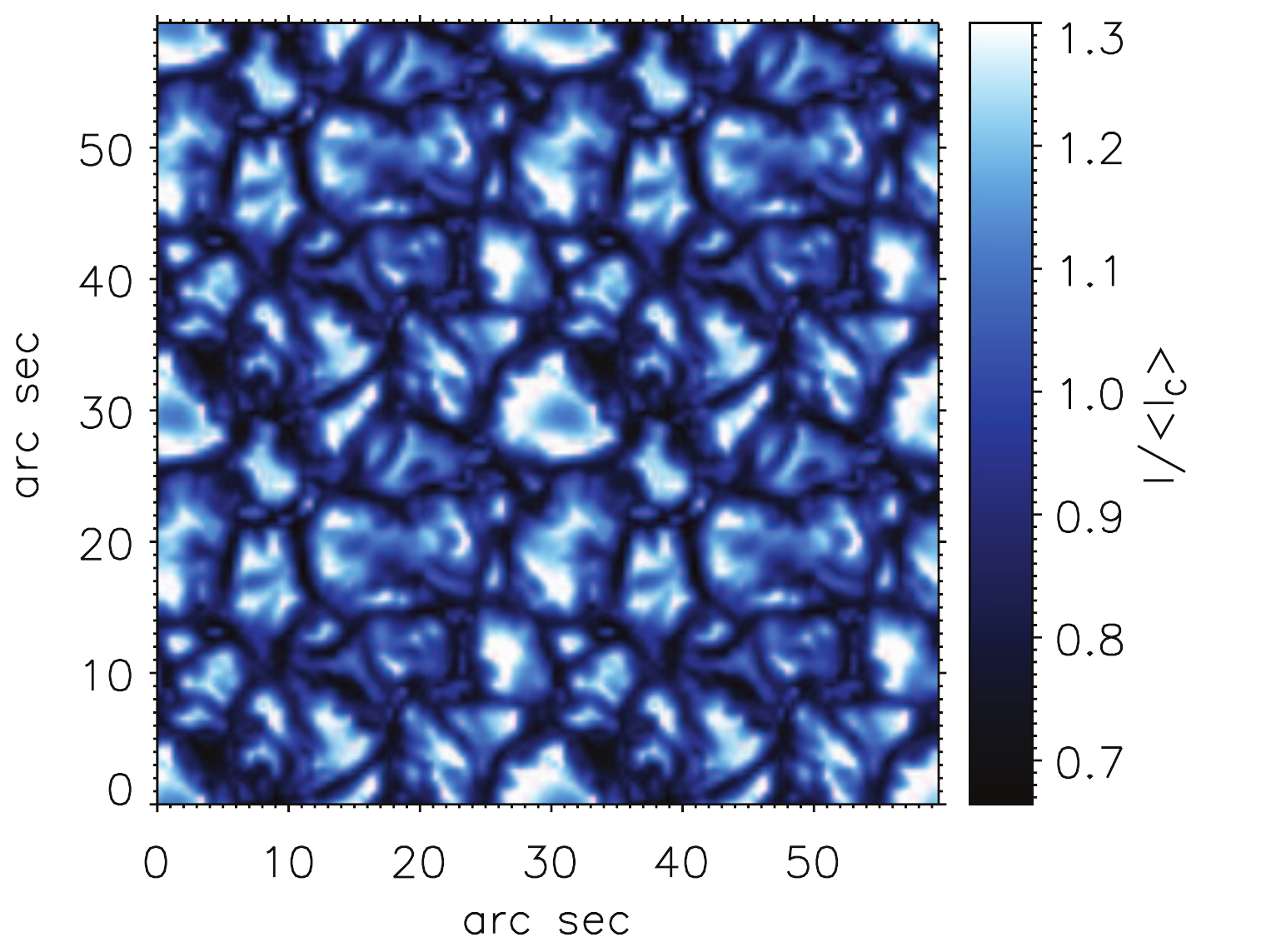}
\includegraphics[width=0.5\textwidth]{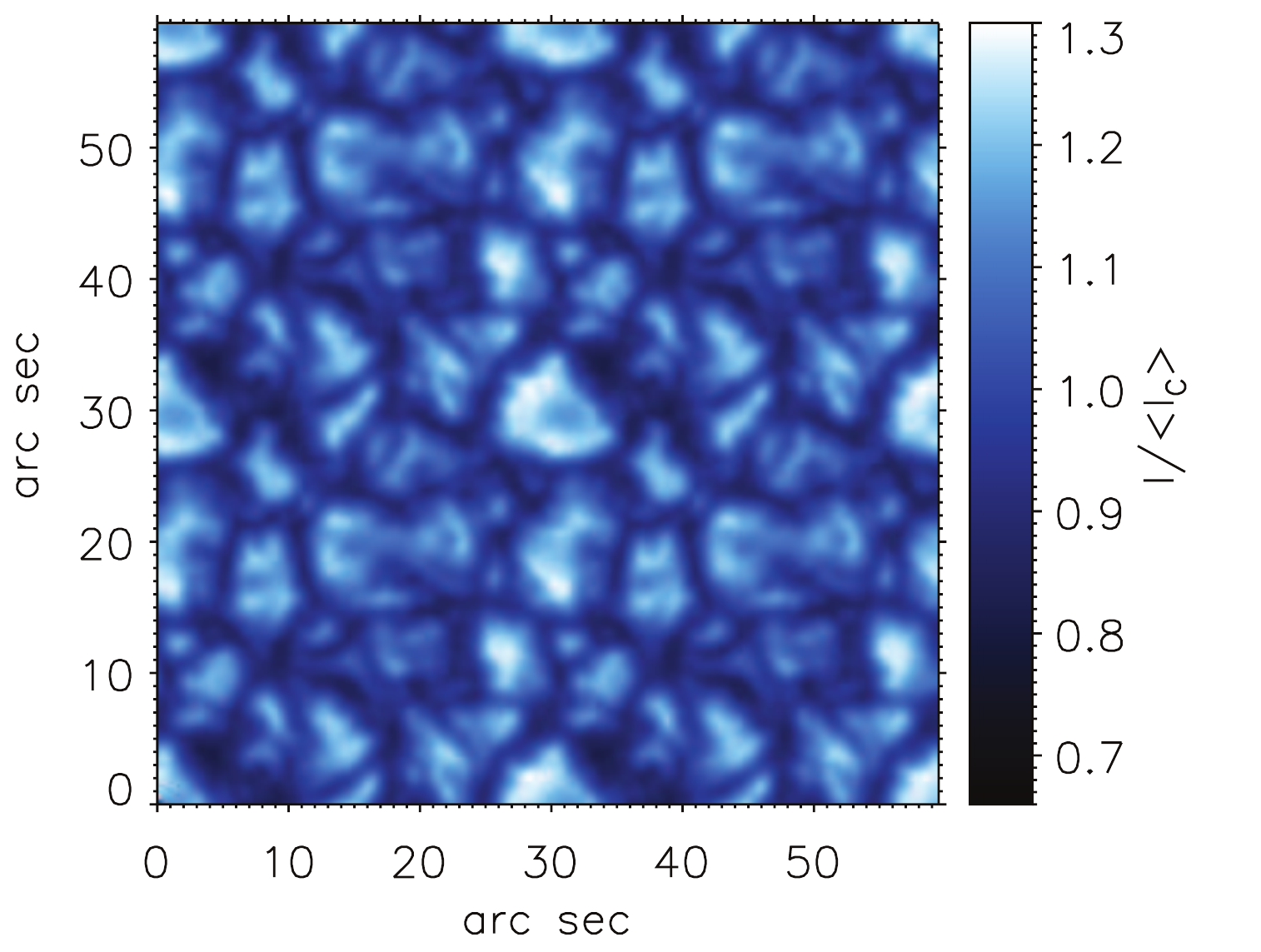}
\end{minipage}
\begin{minipage}[h!]{\textwidth}
\includegraphics[width=0.5\textwidth]{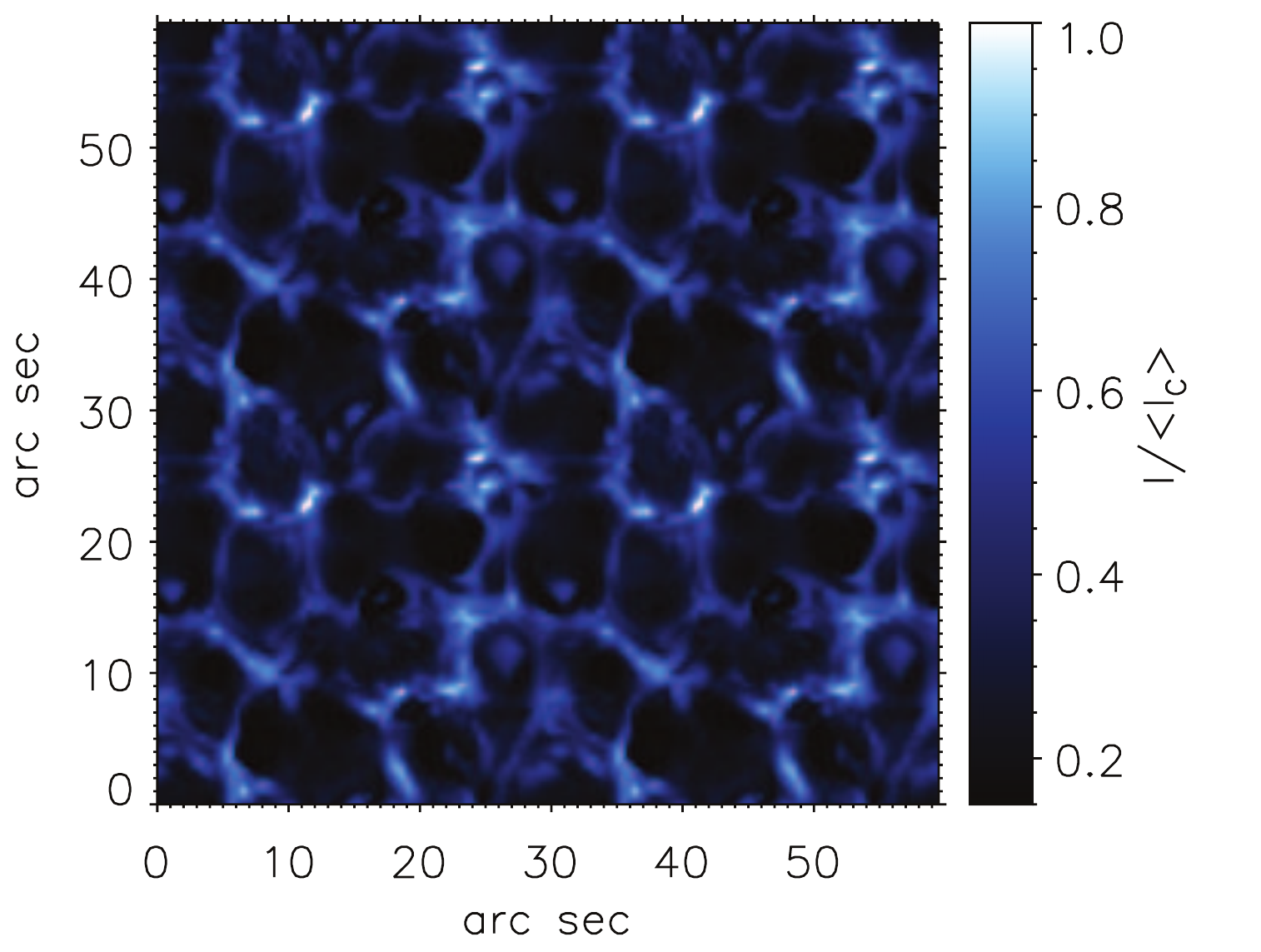}
\includegraphics[width=0.5\textwidth]{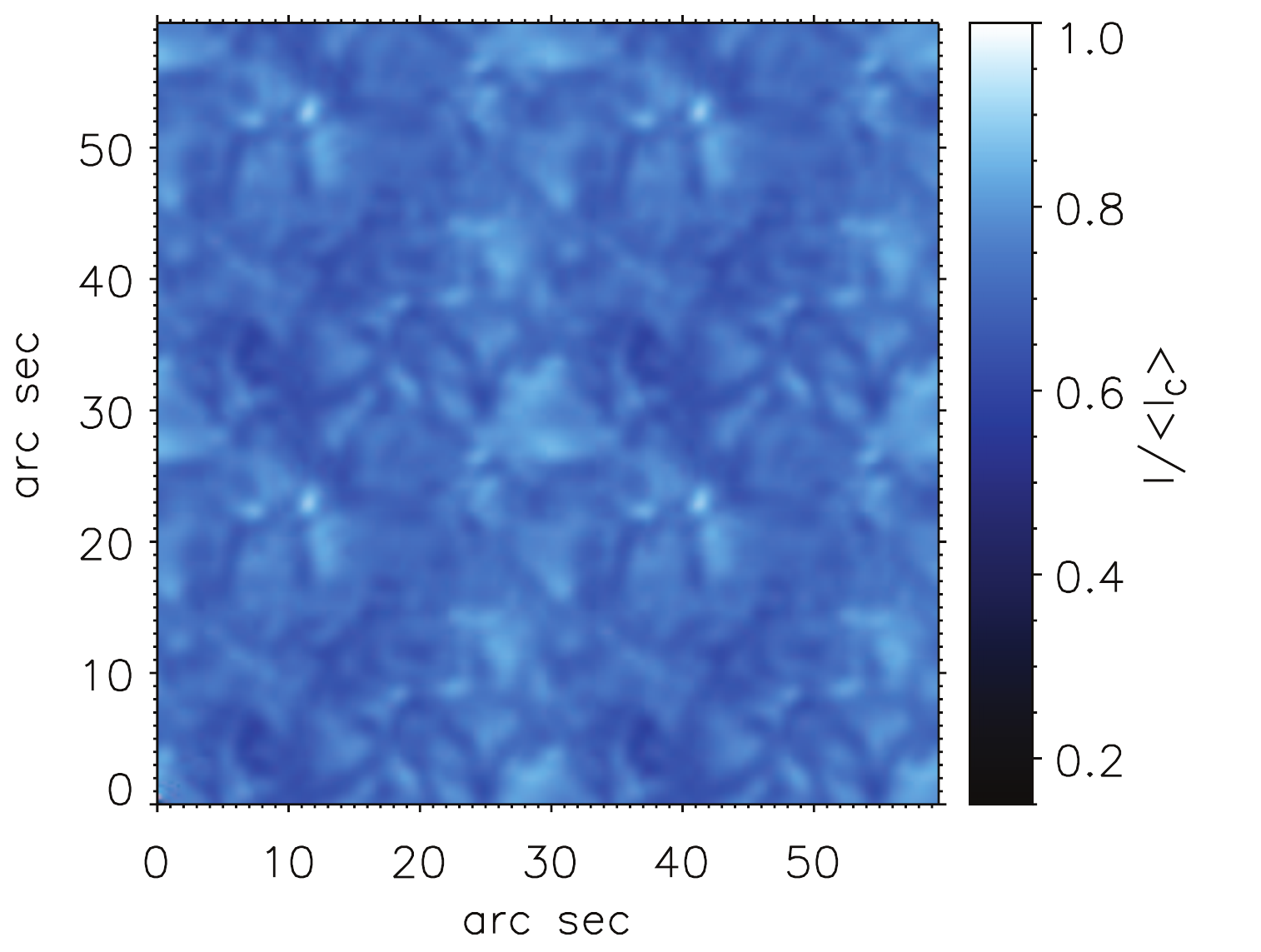}
\end{minipage}
\begin{minipage}[h!]{\textwidth}
\includegraphics[width=0.5\textwidth]{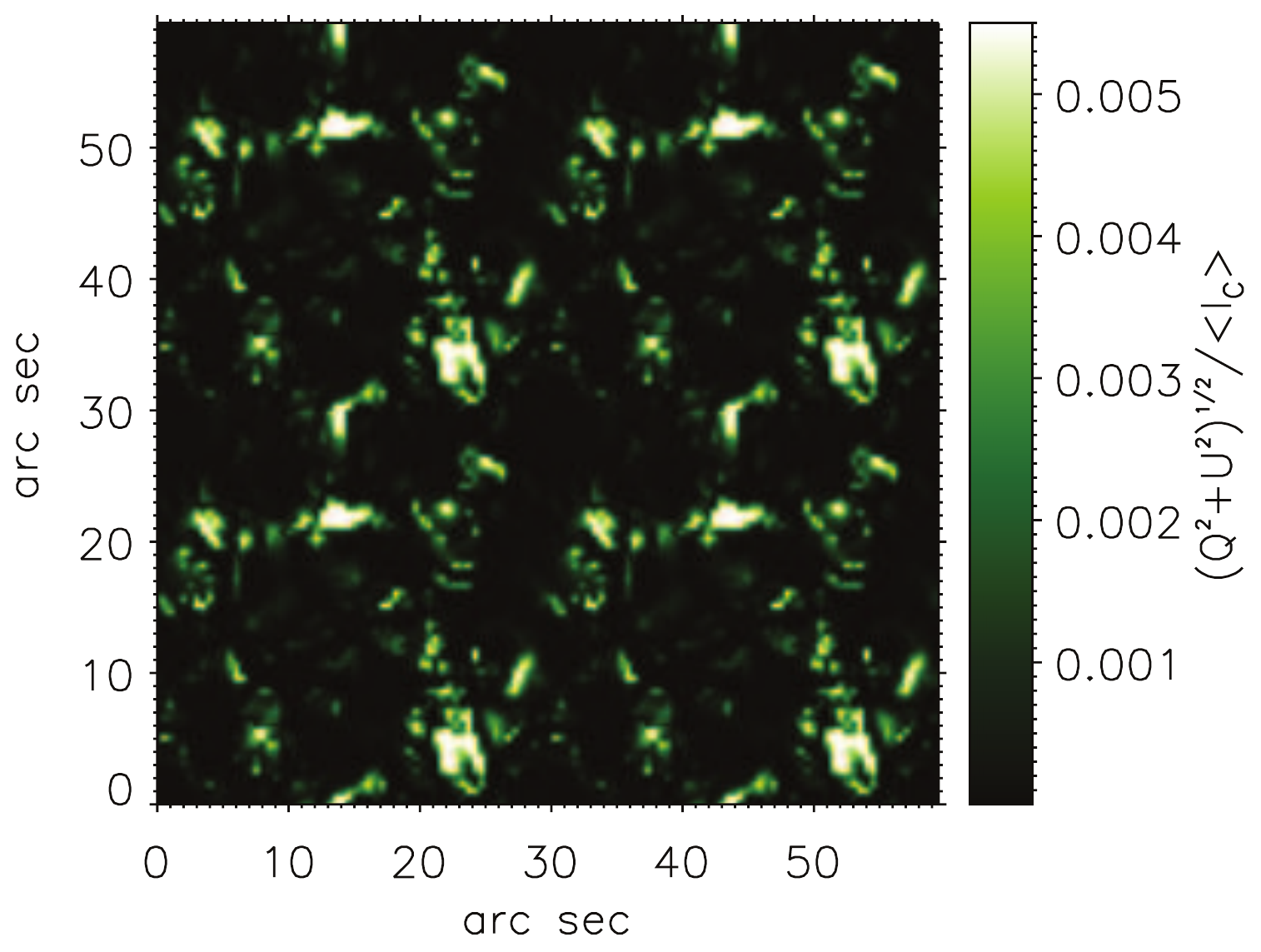}
\includegraphics[width=0.5\textwidth]{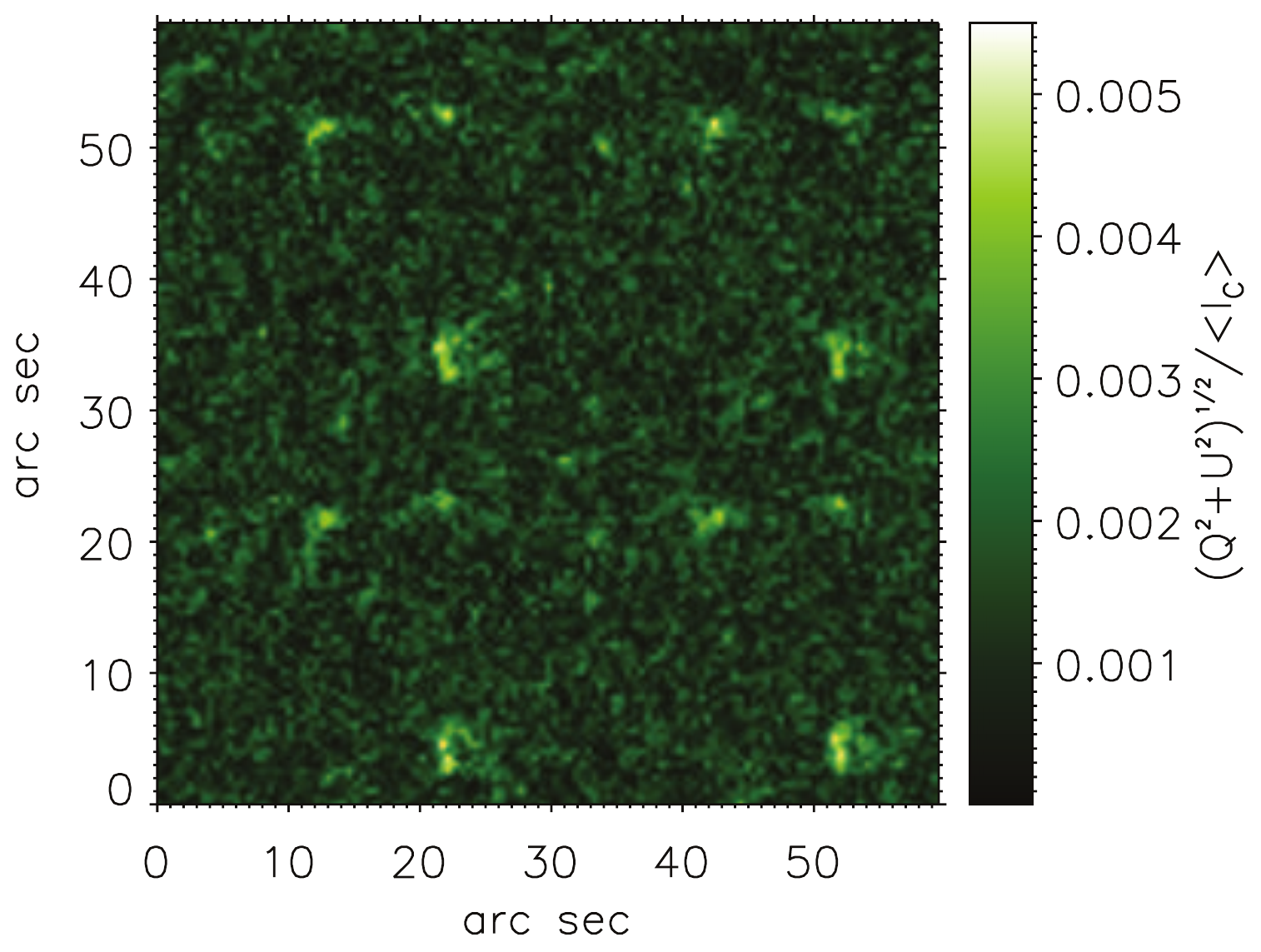}
\end{minipage}
\begin{minipage}[h!]{\textwidth}
\includegraphics[width=0.5\textwidth]{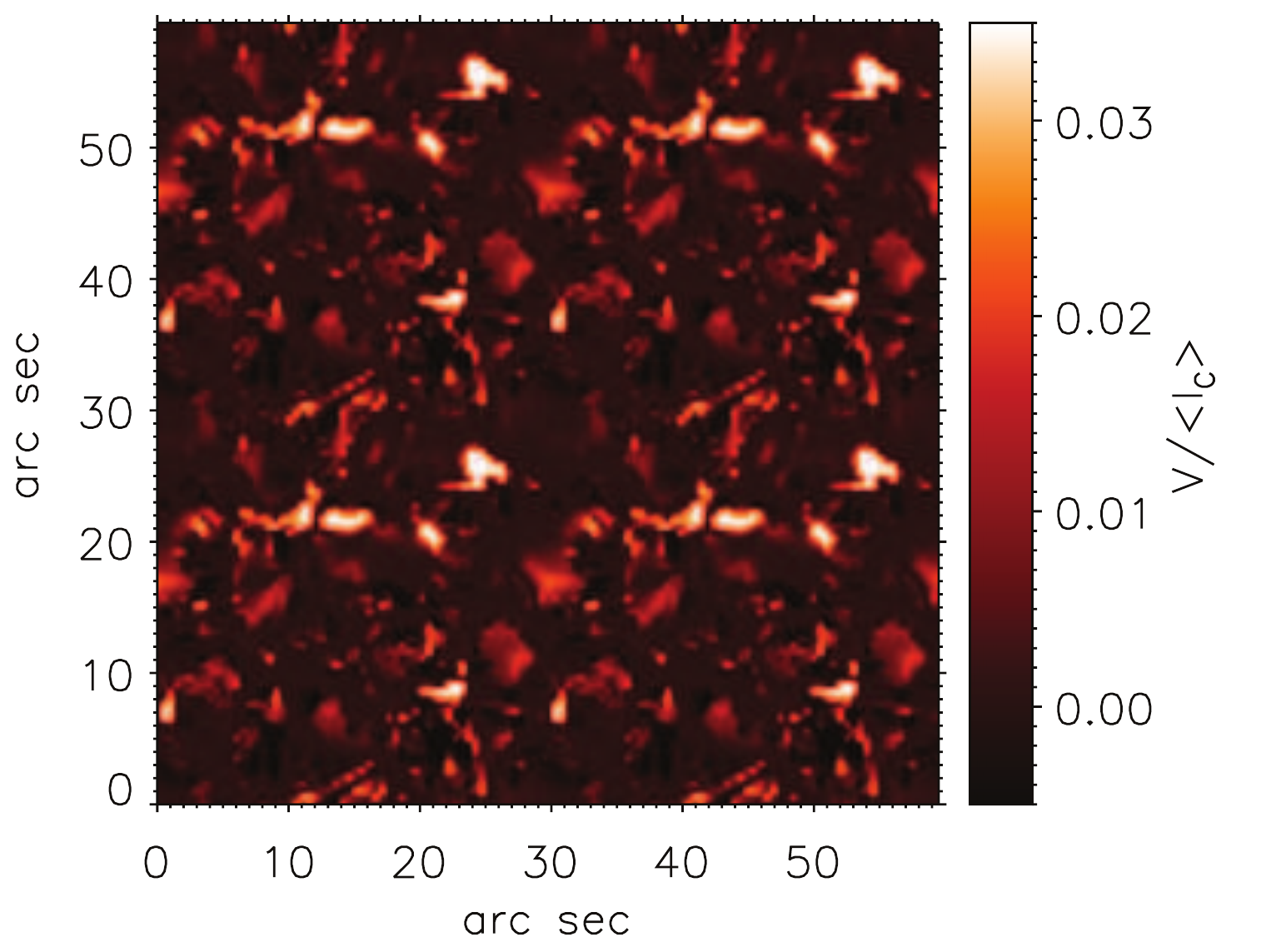}
\includegraphics[width=0.5\textwidth]{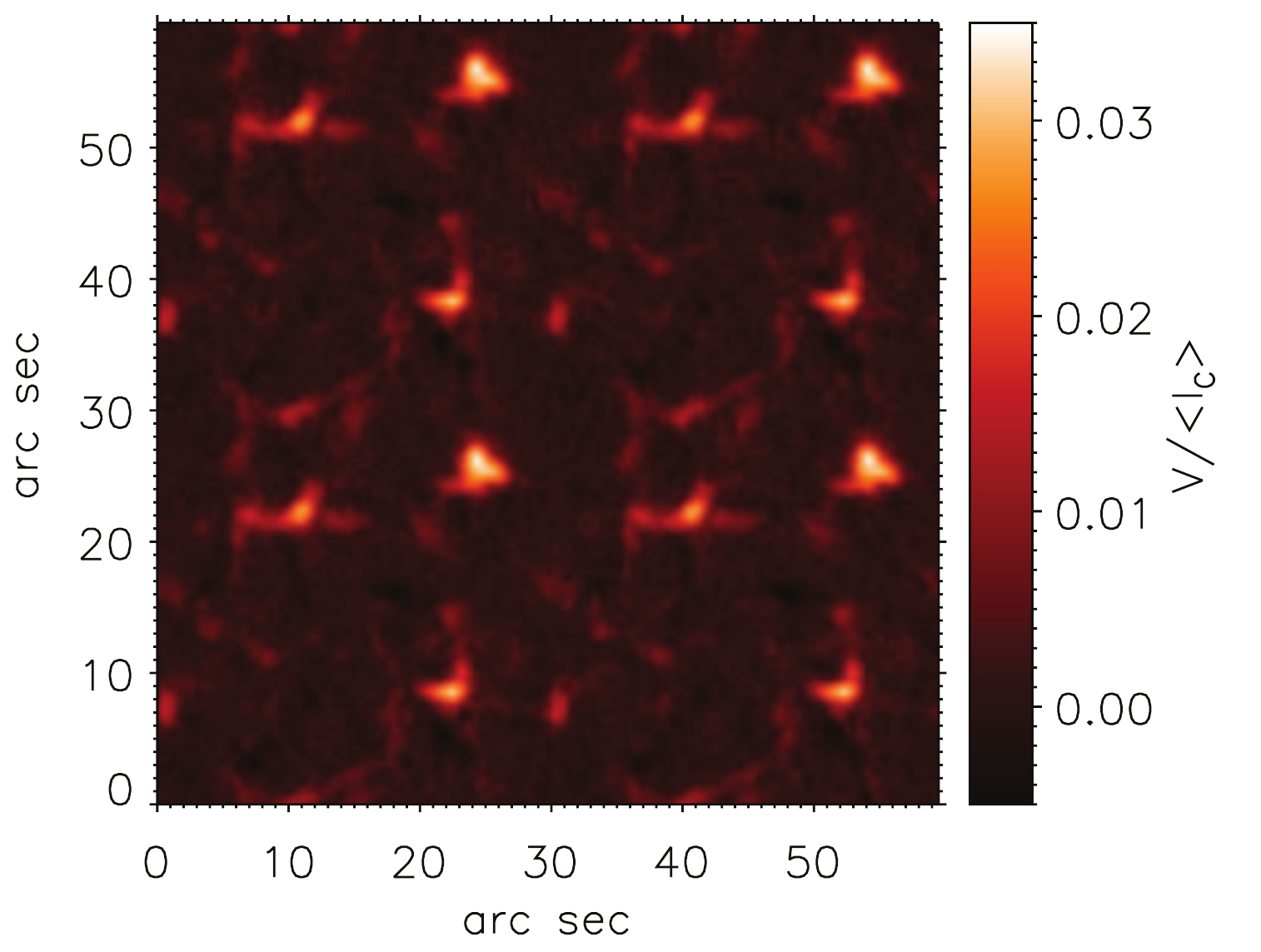}
\end{minipage}
\caption{Example of the influence of SOPHISM on intensity and polarization. The {\it left column} of panels shows the original data from an MHD simulation, the {\it right column} of panels show the results after running SOPHISM. {\it From top to bottom}: Stokes $I$ at $-400$~m\AA \ from line center, Stokes $I$ at line center, $\sqrt{Q^2+U^2}/I$ at $-80$~m\AA \ and $V/I$ at $-80$~m\AA \ from line center. The last two correspond to the degree of linear and circular polarization. Note the change of the colorbars from the original data to the simulation results. The configuration of SOPHISM that was used here and further details about the images are described in the text}
\label{fig:SOPHISM}
\end{figure}

The simulator is programmed in the Interactive Data Language\footnote{IDL is a product of EXELIS Visual Information Solutions, http://www.exelisvis.com/} (IDL) in a modular structure, each module deals with one aspect of the instrument. The modules are mostly independent from each other and can individually be enabled or disabled. The simulation runs can be saved at each step and made available for subsequent calculations. Also, the code is very flexible, with various parameters that can be modified. Presently, the modules and effects covered are the following:
\begin{itemize}
\item Input: this module prepares the input data to be used in the simulation run. If needed, temporal interpolation is performed, as well as spatial operations such as replication of the FOV (for simulation data with periodic boundary conditions) or, considering the Sun-spacecraft distance, scaling from the original spatial resolution to that of the detector.
\item Jitter: this module represents the vibrations induced by the spacecraft, including also the correction by the ISS. A random shift of the FOV is generated and then filtered in frequency according to different possibilities, including a jitter model similar to the {\it Hinode} spacecraft \citep{Katsukawa2010}. Next, the resulting jitter is diminished by means of the ISS attenuation function.
\item Polarization: the polarization modulation of the incoming light is parametrized in this module. This comprises the Mueller matrix of the system and the liquid crystal variable retarders (LCVRs) settings, such as orientation and retardances, to produce the desired modulation of the data.
\item Spectral Profile: the spectral transmission profile of the instrument is calculated here at the user-defined wavelength positions. The resulting transmission curve, a combination of the prefilter and etalon transmissions, is convolved with the data.
An example of a simulated spectral transmission profile can be seen in Figure~\ref{fig:filtergraph}.
\item Optical Aberrations: the MTF of the system with its different aberrations (e.g. defocus, astigmatism, coma,...) is characterized and convolved with the data.
\item Pupil apodization: since the etalon is placed in a telecentric mounting, close to focus, the light is converging over it and enters as a cone. This produces the so-called pupil apodization~\citep{1998A&AS..129..191B,2000A&AS..146..499V}, which results in a radial gradient of intensity and phase over the pupil. This effect has consequences for the spatial resolution and spectral transmission which are taken into account in this module.
\item Focal Plane: all the detector effects are included in this module (e.g. dark current, readout noise, flat-field, shutter, etc.).
\item Accumulation: a given number of exposures are added directly on the instrument in order to achieve a better signal-to-noise ratio.
\item Demodulation: the Stokes vector is recovered here using the demodulation matrices calculated at the polarization module.
\item Inversion: the MILOS inversion code~\citep{2007A&A...462.1137O}, based on a Milne-Eddington approximation, is used to retrieve the full vector magnetic field and LOS velocities maps from the demodulated Stokes vector obtained by the instrument.
\end{itemize}
Figure~\ref{fig:SOPHISM} shows an example of a simulation run. Although in the following part of this paper we only use Stokes $I$, we present here results for the magnetic Stokes parameters as well. The left panels correspond to the input data~\citep[MURaM MHD simulations with an average magnetic field of 50 G,][]{2005A&A...429..335V}, the right panels to the output of SOPHISM. The images presented here are intensity at line center and in the continuum, and $\sqrt{Q^2+U^2}/I$ and $V/I$ at $-80$~m\AA \ in the line wing, corresponding to the degree of linear and circular polarization, respectively. The simulations represent an observation at disk center from perihelion ($0.28$~AU) and have a size of $12 \times 12$~Mm (the original simulations have a size of $6\times 6$ Mm, but are replicated in the horizontal spatial dimensions in order to increase the FOV, taking advantage of its periodic boundary conditions). The simulations presented in this section have the following main characteristics: jitter with an 
RMS normalized to $0.5''$ and attenuated by the ISS; polarization modulation with identity Mueller matrix and ideal modulation; a prefilter with a FWHM of 3 \AA\ and the etalon at the six positions given in Sect.~\ref{sect:PHI} with a resulting transmission FWHM of $\sim$ 90 m\AA\ along with pupil apodization considerations; an aberrated wavefront normalized to $\lambda$/10; dark current and photon and readout noises; and 12 accumulations.

In the intensity images presented in Figure~\ref{fig:SOPHISM}, the most evident effect of the simulated instrument is a loss of spatial and spectral resolution. Because of the lower signal of the magnetic Stokes parameters $Q$, $U$, and $\vlos$, other effects become apparent in the linear and circular polarization images, mainly crosstalk from Stokes $I$ due to spacecraft jitter and noise components. The random parts of the simulation, like the jitter generation, noises, etc., are produced in every run. So, a different simulation run, even with the same settings, will yield different results. This is especially noticeable with the jitter crosstalk in the $Q$, $U$, $\vlos$ parameters, which may show different contributions in the same parameter at the same spectral position because of larger or smaller jitter shifts coinciding in the same data product.

The output of SOPHISM depends significantly on the initial settings mentioned above. Although the configuration used for the results shown in Figure~\ref{fig:SOPHISM} corresponds to the present assumptions for the behavior of PHI, these simulations will not be identical with the actual instrument. The default settings and characteristics of most elements are taken from other instruments or calculated from a theoretical point of view. For example, the frequency filter for the spacecraft jitter is taken from {\it Hinode} and the prefilter curve is computed theoretically. As the design of PHI proceeds, more precise settings can be used in the simulator, including parameters derived from measurements in the lab. The development of the code will continue during the following years to represent further realistic characteristics of the instrument, such as temperature dependencies, aging (radiation effects, inefficiencies,...), FOV dependent Mueller matrices, etc.

\section{Simulating PHI Time-Series for Helioseismology}
\label{sect:synth_data}

We are now at the stage where a detailed strategy has to be developed to maximize the helioseismology output of the mission given the various limitations imposed by the mission (e.g., challenging orbit, and an expected telemetry allocation of 51 Gbit per orbit). In this section we present synthetic data with the same properties as expected from the High Resolution Telescope of the PHI instrument on {\it Solar Orbiter} and begin characterizing the properties of the data for helioseismic studies (e.g. the expected power spectra). Starting from realistic radiative MHD simulations computed with the Stagger code~\citep{2000SoPh..192...91S}, computing line profiles with the SPINOR code~\citep{2000A&A...358.1109F}, and simulating the PHI instrument using SOPHISM (see Sect.~\ref{sect:SOPHISM}), we have generated a time-series of synthetic Dopplergrams. These Dopplergrams are models for what should be available onboard the {\it Solar Orbiter} satellite.

\subsection{Steps in the Generation of Synthetic Data}
\label{sect:steps}

\subsubsection{Simulations of Solar Convection}
We start from simulations of solar surface convection for the quiet Sun computed with the Stagger code. These simulations exhibit solar oscillations and have previously been used in helioseismic studies, mostly for analyzing the excitation mechanism of solar oscillations~\citep[e.g.][]{2001ApJ...546..585S, 2001ApJ...546..576N, 2003A&A...403..303S, 2004SoPh..220..229S} but also for testing methods used in helioseismology~\citep[e.g.][]{2007ApJ...659..848Z, 2007ApJ...657.1157G, 2007ApJ...669.1395B, 2009SoPh..257..217C} or for modeling helioseismic observations~\citep[e.g.][]{2012ApJ...760L...1B}.

We use a simulation run with a size of $96 \times 96 \times 20$~Mm, corresponding to $2016 \times 2016 \times 500$ grid points. The horizontal resolution is constant (about 48 km) and the vertical resolution varies with height between 12 and 79 km. We analyze 359 minutes of the simulation, for which we have snapshots of the entire state of the system with a cadence of one minute corresponding to the planned cadence of PHI. We assume in this work that variations on shorter timescales have a negligible effect. We plan to test this assumption using a short run of the simulations with a high cadence.

In order to reduce computation time we analyze a $48 \times 48$~Mm sub-domain from the simulations. This is a small patch of the field of view of the HRT of PHI ($16.8'$, corresponding to $\sim 200$~Mm at perihelion) but it is sufficient for studying the solar oscillations in the synthetic data.

\subsubsection{Computation of Line Profiles}
We synthesize line profiles for the Fe I 6173 \AA \ line for every single pixel of the simulations using the SPINOR code with atomic parameters taken from the Kurucz atomic database~\citep[$\log gf = -2.880$,][]{1988JPCRD..17S....F} and the iron abundance ($\textmd{A}_{\rm Fe} = 7.43$) from~\citet{2002A&A...391..331B}. Note that SPINOR allows simulations of observations at heliocentric angles $\rho > 0$ by a synthesis of spectra obtained from inclined ray paths.

\subsubsection{Using SOPHISM}
Since we want to analyze a relatively long time-series, we configure SOPHISM in a way that minimizes the computation time while still covering all relevant effects for helioseismology. One of the computationally most demanding parts of SOPHISM is the polarimetry. Polarimetry is not of great importance for helioseismology, so we do not include it in the following. We further reduce the computation time by decreasing the number of simulated exposures. When scanning the Fe I 6173 \AA \ line, PHI will make several exposures for each of the wavelength positions over a total time of $60$~s. We assume observations at six wavelength positions, located at $-400$, $-160$, $-80$, $0$, $80$, and $160$~m\AA \ relative to the line center. During this time, the Sun evolves, causing small differences between the individual exposures. This means that the order in which the wavelengths are observed affects the resulting image. The Stagger simulation data we use are sampled at a cadence of $60$~s, so we can compute only one 
exposure. The number of exposures and the order in which they are taken are also important for the impact of spacecraft jitter and photon noise. So, in order to get realistic results when computing only one exposure, in this section we treat the jitter and photon noise in a slightly different way than the standard SOPHISM.


\subsubsection{Spacecraft Jitter}
\begin{figure}
\begin{minipage}[h!]{\textwidth}
\includegraphics[width=0.521\textwidth]{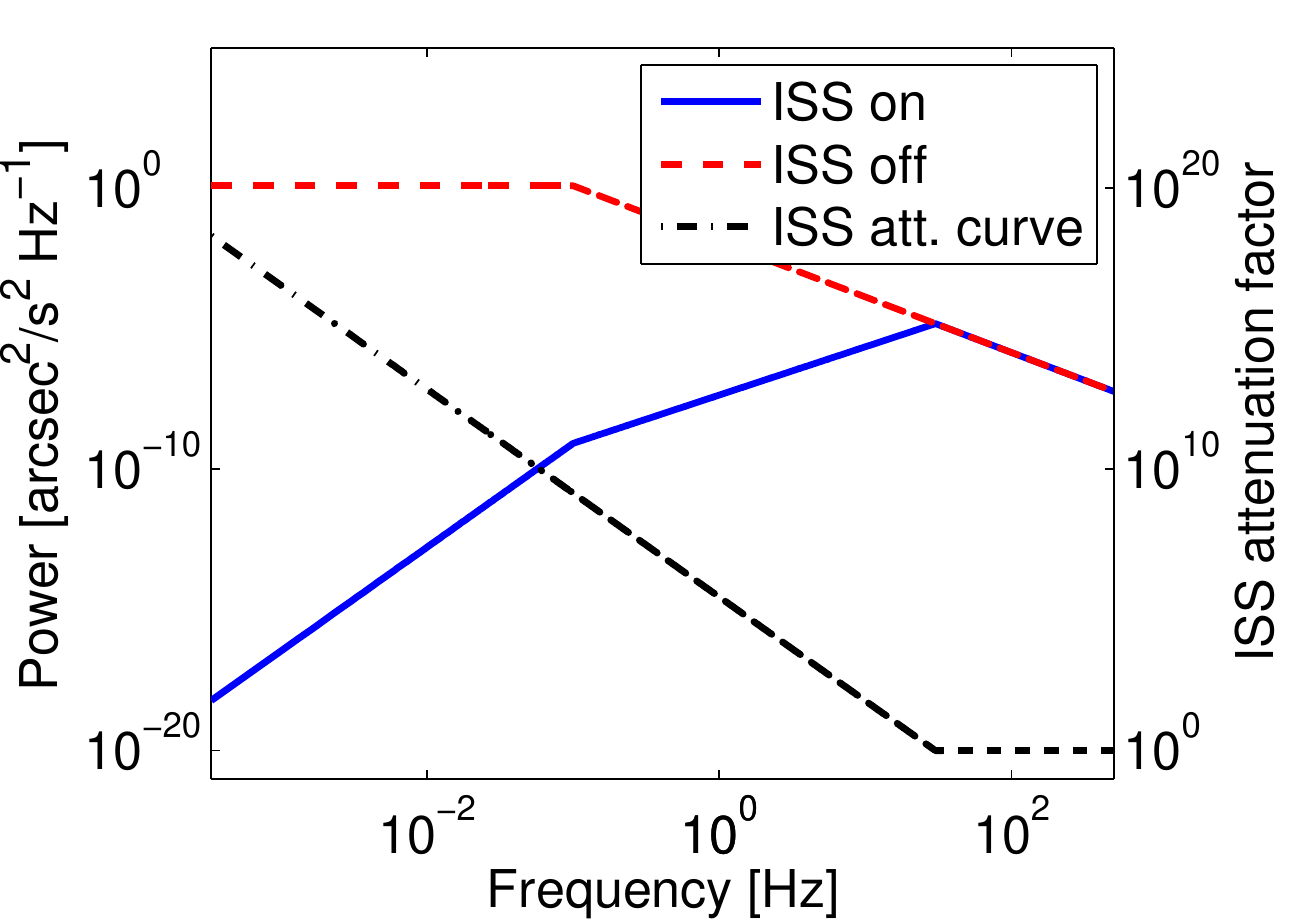}
\includegraphics[width=0.479\textwidth]{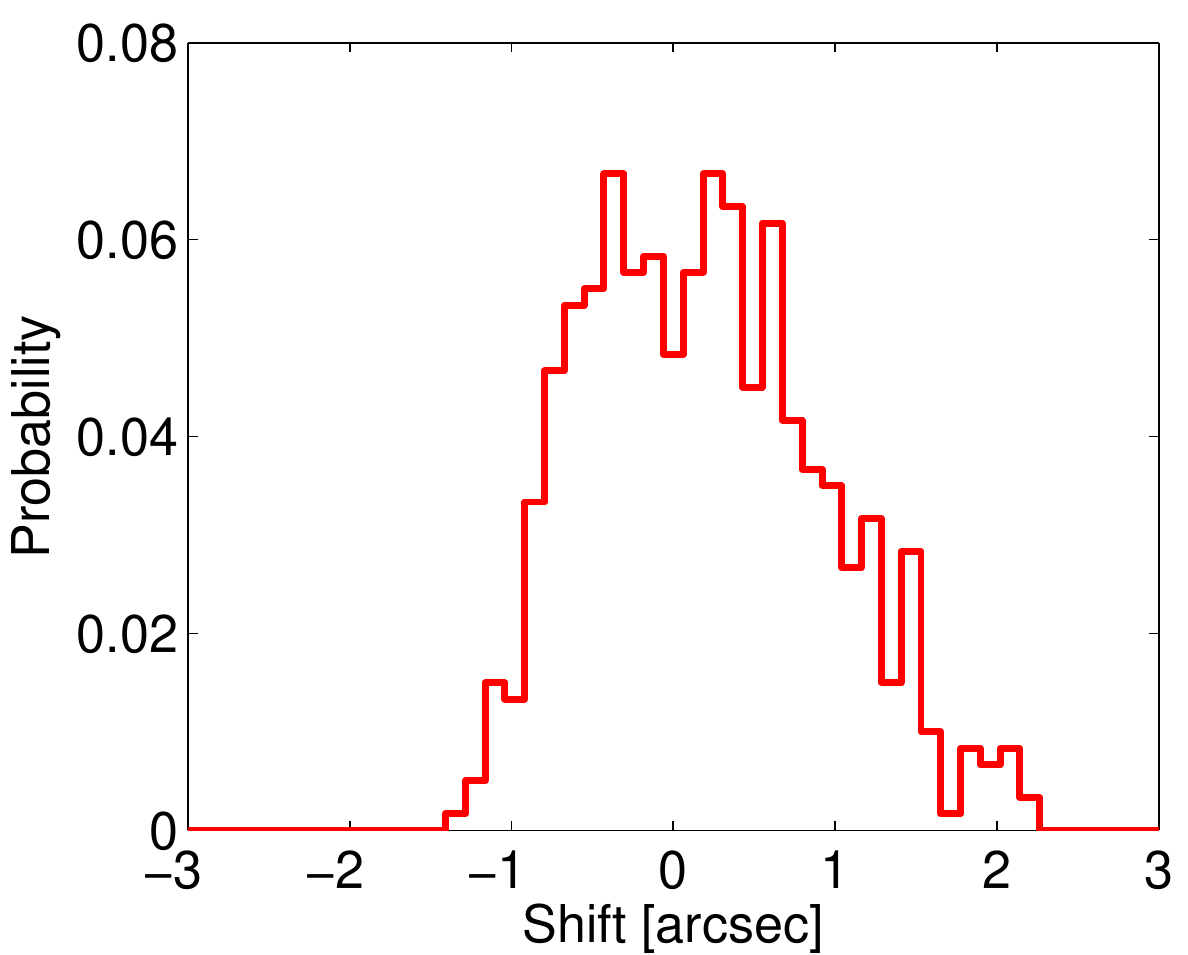}
\end{minipage}
\caption{{\it Left}: Power spectral density of the shift introduced by the modeled spacecraft jitter. We assume the jitter to be similar to that of the {\it Hinode} spacecraft and to follow the spacecraft requirement  (see text for details). When the ISS is turned off ({\it dashed curve}), we use a power law to model the power spectral density of the {\it Hinode} jitter at frequencies above $0.1$~Hz ($P(\nu)\propto \nu^{-1.98}$) and a constant power spectral density for frequencies smaller than $0.1$~Hz. We model the influence of the ISS by dividing the power spectral density of the jitter ({\it dashed curve}) by a model for the attenuation curve of the ISS ({\it dash-dotted curve}). When the ISS is turned on, the jitter is reduced significantly ({\it solid curve}). The ISS works most efficiently at low frequencies. {\it Right}:  Example of a histogram of the shifts introduced by the jitter (ISS turned off). We used a time-series with a duration of 60~s, which corresponds to the cadence of PHI. The 
distribution has an RMS of $0.7''$}
\label{fig:jitter}
\end{figure}

The exact behavior of the spacecraft jitter of {\it Solar Orbiter} is not known yet. When modeling the jitter, we follow the spacecraft requirement that the RMS of the jitter with frequencies above $0.1$~Hz must not be higher than $0.5''$ (or $0.1''$ when the ISS is turned on). We use a model for the jitter that depends on frequency with a power spectral density similar to that of the {\it Hinode} spacecraft. For frequencies higher than $0.1$~Hz, we model the power spectral density with a power law fit of the {\it Hinode} jitter curve (see left part of Figure~\ref{fig:jitter}) normalized to an RMS of $0.5''$. We are not aware of specifications for the jitter at lower frequencies. Here we assume a constant power spectral density of the jitter, corresponding to the power at $\nu=0.1$ Hz. This low frequency part increases the RMS from $0.5''$ to $0.7''$. We also model a jitter curve when the ISS is activated by dividing the power spectral density of the jitter by a modeled attenuation curve of the ISS (black 
curve in Figure~\ref{fig:jitter}). The attenuation is strongest at low frequencies and reduces the RMS from $0.7''$ to $0.03''$, which is much less than the pixel size ($0.5''$). We express the jitter in the Fourier domain by using a fixed amplitude for each frequency (derived from the power spectral density) and a random phase. We generate a time-series of the jitter by taking the inverse Fourier transformation.

The influence of the jitter depends on the order in which the filtergrams are measured. We assume the jitter to act on all wavelengths in the same manner by using the same time-series for all wavelengths. Since we compute only one exposure, we do not model the jitter by shifting individual images. Instead, we convolve the image with a distribution corresponding to the accumulated shifts introduced by the jitter within the PHI cadence. The right part of Figure~\ref{fig:jitter} shows an example for this distribution. Since the jitter is derived from {\it Hinode}, it does not correspond to the jitter of the final {\it Solar Orbiter} spacecraft. However, as will be shown in Sect.~\ref{sect:power}, the influence of the jitter on the power spectrum is very small. Hence, we do not expect that our conclusions are affected by selecting a specific model for the jitter.

\subsubsection{Point Spread Function}
\label{sect:Optics}
\begin{figure}
\begin{minipage}[h!]{\textwidth}
\includegraphics[width=0.4978\textwidth]{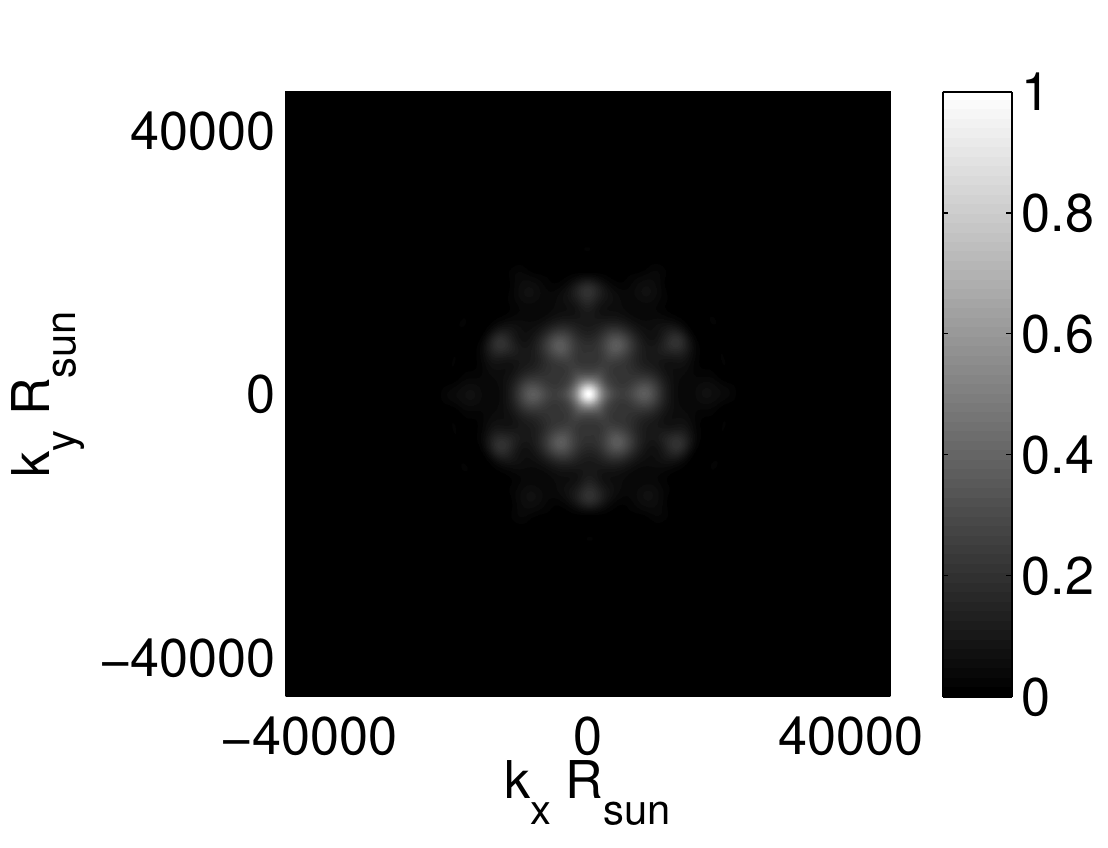}
\includegraphics[width=0.5022\textwidth]{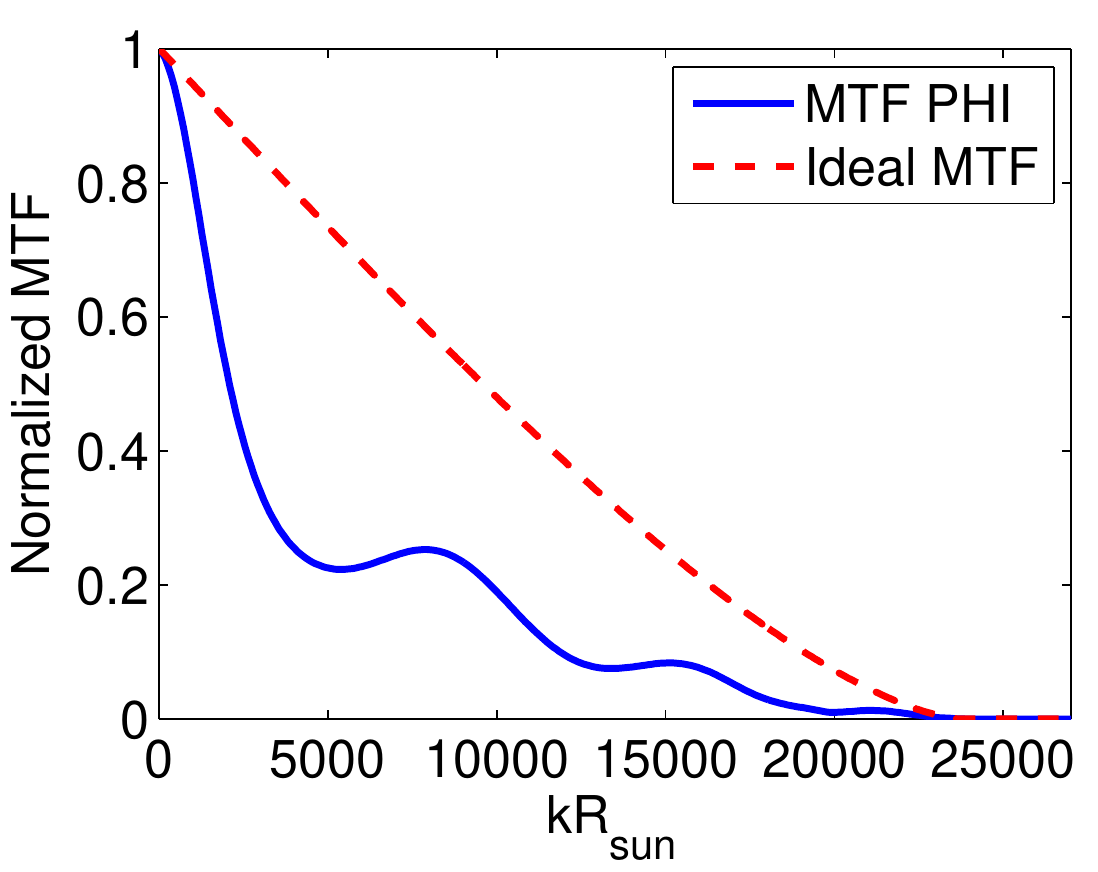}
\end{minipage}
\caption{Modulation Transfer Functions used in this study as a function of wavenumber when observing at perihelion ($0.28$~AU). {\it Left}: MTF computed from a modeled instrument, where the entrance window is not designed in the best possible way, {\it Right}: Azimuthal average of this MTF ({\it solid curve}) and an MTF corresponding to a diffraction-limited instrument ({\it dashed curve})}
\label{fig:MTF}
\end{figure}

An important characteristic of PHI is the Modulation Transfer Function (MTF). Since the MTF is mostly determined by the optical properties of the HREW when it is exposed to the environmental conditions in the orbit, the final MTF is still subject to tests and simulations. For this work, we use two different Modulation Transfer Functions corresponding to the best and the worst possible outcome. The first MTF represents a diffraction-limited instrument, we assume the Point Spread Function (PSF) to be an Airy function corresponding to the aperture of the High Resolution Telescope (14 cm). The second one is an MTF that was computed from a model of the PHI instrument, where the entrance window was located at a suboptimal position within the heat shield of the spacecraft\footnote{Recently, preliminary results from a study for the optimum HREW location have shown that a nearly diffraction-limited telescope under all orbit conditions is feasible.} (see Figure~\ref{fig:MTF}). Compared to 
the MTF for the diffraction-limited instrument, the MTF computed from the model of PHI strongly decreases with the wavenumber and is also very asymmetric. 

\subsubsection{Photon Noise}
The last step is modeling the detector. Here we resample the data to the detector plate scale ($0.5''$ per pixel, corresponding to $\sim 100$~km at perihelion) and add photon noise to the data.

PHI will accumulate several exposures to generate one final image in order to increase the signal-to-noise ratio. The requirement for PHI for polarimetry is an accuracy of $\pm 10$~G per pixel for determining the LOS magnetic field, $\pm 200$~G for the transverse component of the magnetic field and $\pm 15$~m/s for the LOS velocity. This demands an instrument requirement for the signal-to-noise ratio of the mean continuum of at least 1000. Here we assume the noise to follow the requirement.

Since we simulate only one image, representing the accumulation of many individual exposures, we can add the noise directly to the intensity images using a normal distribution with dispersion
\begin{equation}
\sigma (\lambda,x,y) = 10^{-3} \sqrt{ I(\lambda,x,y) \left < I_{\rm c} \right >}.
\end{equation}
Here, $I(\lambda,x,y)$ is the intensity at a given wavelength and a given pixel and $\left < I_{\rm c} \right >$ is the spatially averaged continuum intensity at disk center. This results in the desired signal-to-noise ratio:
\begin{equation}
 \frac{I}{\sigma} = 
10^3  \sqrt{ \frac{I(\lambda,x,y)}{\left < I_{\rm c} \right > }} .
\end{equation}
Since PHI will perform an onboard correction of the raw images for darks and flat-fields, we do not model their influence here. We also neglect quantization noise. The raw images provided by the detector will have a size of $14$~bits per pixel, so the influence of quantization noise is far below the photon noise level.
Another source of noise is caused by variations of the exposure time. The requirement for this noise is 7 m/s, demanding a target accuracy of the exposure time of 450 ppm~\citep{shutter}.


\subsection{Synthetic Intensity and Velocity Maps}
Applying the individual steps described in Sect.~\ref{sect:steps} results in a time-series of intensity images at different wavelengths as will be available for onboard processing on {\it Solar Orbiter}. Here we determined the LOS velocities by computing the barycenter of the intensity at the six wavelength positions that we have selected.

\begin{figure}
\includegraphics[width=\textwidth]{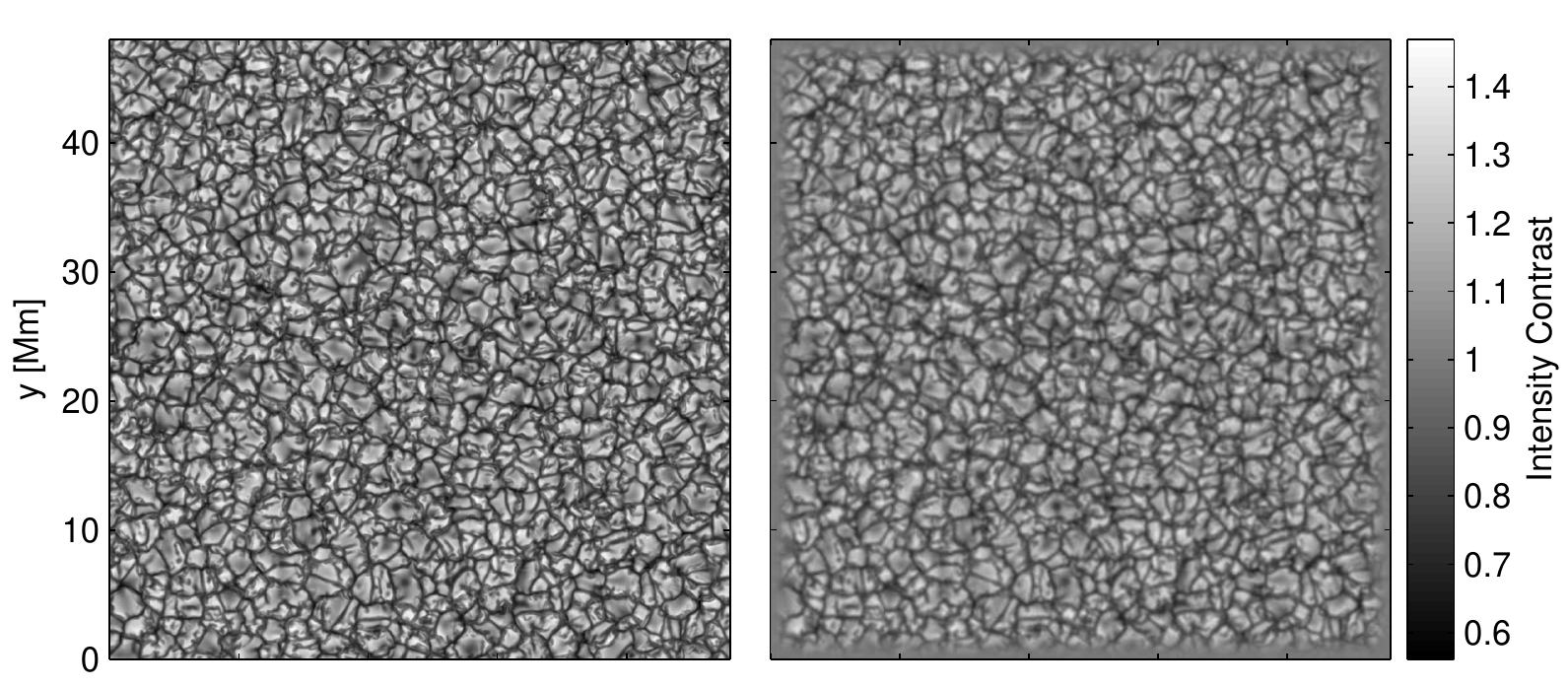}
\includegraphics[width=\textwidth]{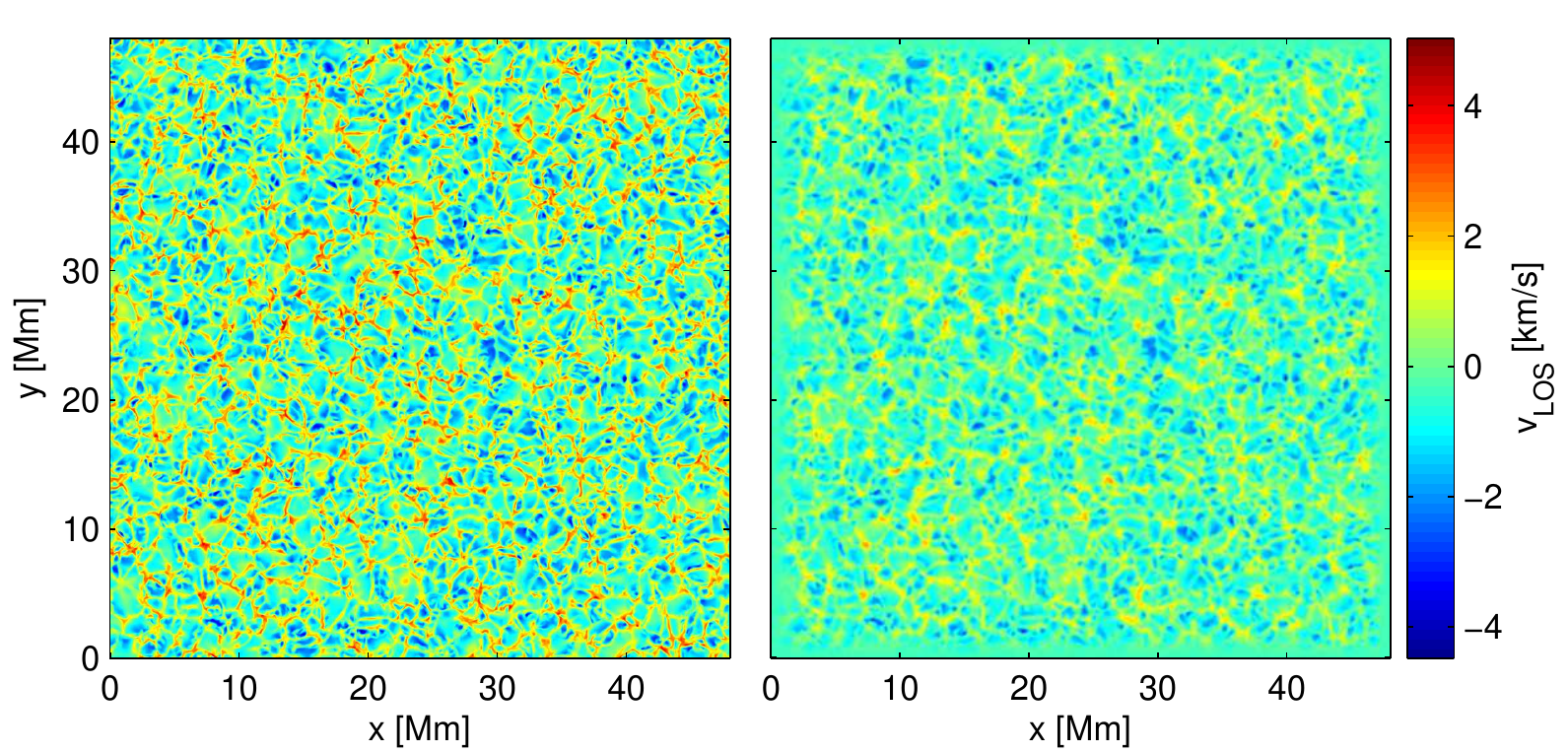}
\caption{Influence of the modeled High Resolution Telescope of PHI on intensity ({\it top}) and line-of-sight velocity ({\it bottom}) when observing disk center from perihelion ($0.28$~AU). We compare data from the line profiles ({\it left}) with our results for the modeled PHI instrument ({\it right}). The dominant signal in these images is granulation. The Dopplergrams clearly show the upflows in the granules and the downflows in the intergranular lanes. In the modeled PHI data, the spatial resolution is reduced due to the MTF. The edges of the SOPHISM data are smooth, because SOPHISM apodizes the image before applying the MTF in order to avoid artifacts in the Fourier spectrum. Here we use an MTF corresponding to a diffraction-limited instrument. The image is only a fraction of the full FOV for the HRT}
\label{fig:intensity_velocity}
\end{figure}

\begin{figure}
\includegraphics[width=\textwidth]{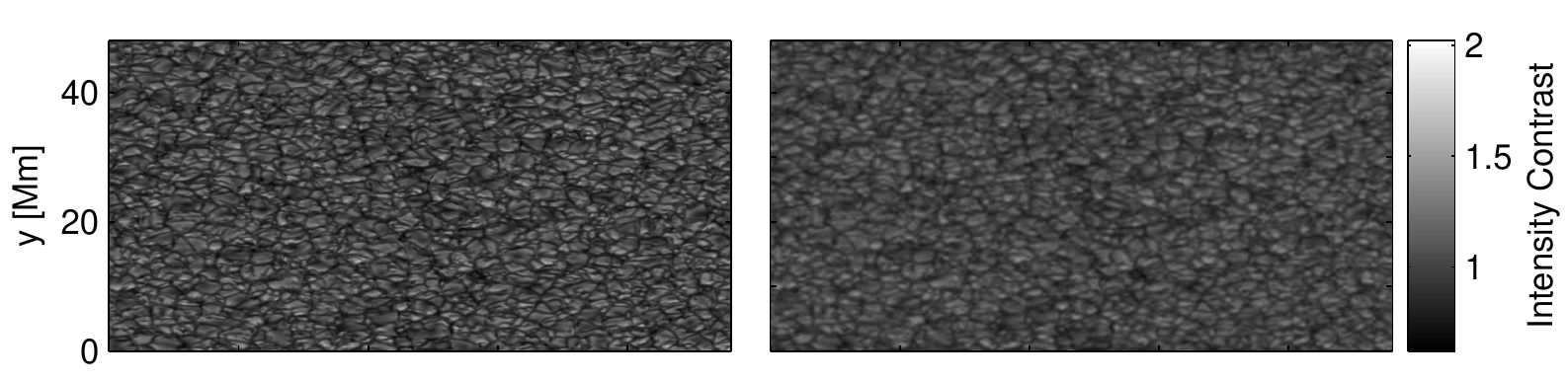}
\includegraphics[width=\textwidth]{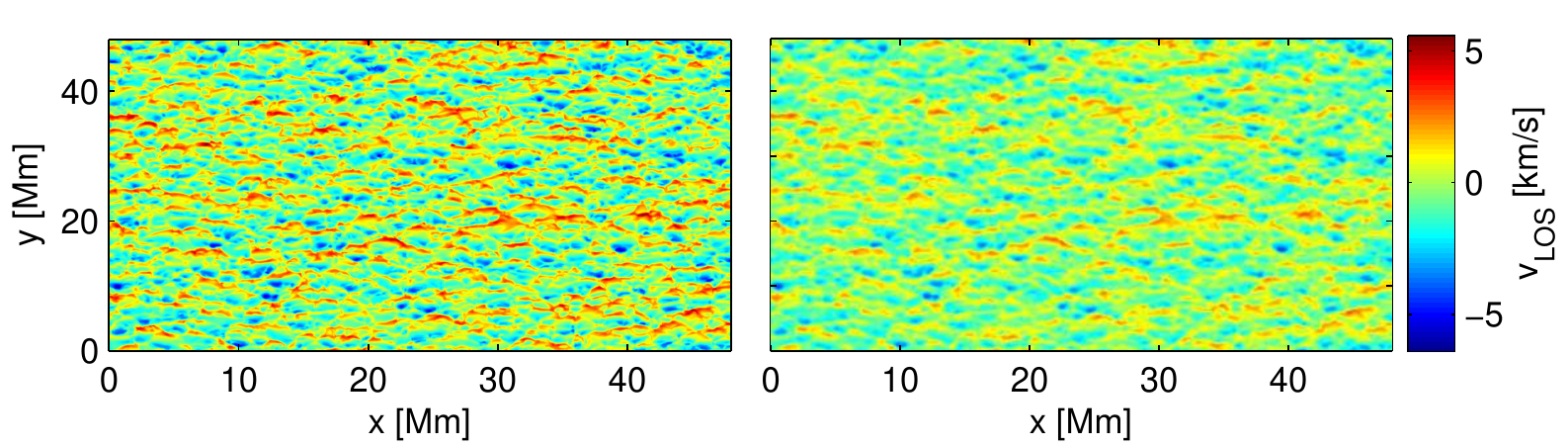}
\caption{Same as Figure~\ref{fig:intensity_velocity} for an observation at $\rho=60^\circ$. Again, we assume an observation from perihelion ($0.28$~AU). Due to foreshortening the images are squeezed in y-direction. The top of the box is directed towards the limb. While the intensity images show granulation, as for the disk center data, the Dopplergrams are now dominated by flows in the horizontal direction, which originate from outflows from the granules}
\label{fig:intensity_velocity_limb}
\end{figure}

We generated two such time-series with a length of 359~min each, one corresponding to an observation of the disk center and another one for an observation closer to the limb (heliocentric angle $\rho=60^\circ$). This corresponds to an observation of the solar poles from a solar latitude of $30^\circ$, as it will be done by {\it Solar Orbiter}. In both cases we have assumed the spacecraft to observe from perihelion ($0.28$~AU). Since the length of the time-series is rather short, we have neglected spacecraft motion. Examples for intensity and velocity maps are shown in Figure~\ref{fig:intensity_velocity} for disk center and in Figure~\ref{fig:intensity_velocity_limb} for $\rho = 60^\circ$.

At disk center, granulation is clearly visible in the intensity images and Dopplergrams. When observing at $\rho=60^\circ$, the granulation can still be resolved  in the intensity images but the contrast between granules and intergranular lanes is reduced. The Dopplergrams at the limb are dominated by horizontal outflows from the granules to the intergranular lanes. After running SOPHISM, the contrast of both the intensity and the velocity is reduced significantly (see Table~\ref{tab:int_v_rms}) due to the MTF. There are strong differences between the two MTFs that we use here, the worst-case MTF reduces the RMS both for intensity and velocity by about a factor of two.

\begin{table}
\caption{RMS for continuum intensity contrast and LOS velocity both for disk center and $\rho = 60^\circ$. We compare data from SPINOR with our results from SOPHISM. Both the MTF corresponding to a diffraction-limited instrument and a worst-case MTF are considered}
\label{tab:int_v_rms}       
\begin{tabular}{lcccc}
\hline\noalign{\smallskip}
& \multicolumn{2}{c}{RMS of intensity contrast} & \multicolumn{2}{c}{RMS of LOS velocity [km/s]}\\
 & Disk center & $\rho=60^\circ$& Disk center & $\rho=60^\circ$\\
 \noalign{\smallskip}\hline\noalign{\smallskip}
SPINOR & $0.17$ & $0.14$ &$1.20$ & $1.54$\\
SOPHISM (diff.-limited MTF) & $0.12$ &  $0.10$ &$0.72$ & $1.08$\\
SOPHISM (worst-case MTF) & $0.06$ & $0.05$ & $0.37$ & $0.62$\\
\noalign{\smallskip}\hline
\end{tabular}
\end{table}

\subsection{Oscillation Power Spectra}\label{sect:power}
\begin{figure}
\includegraphics[width=\textwidth]{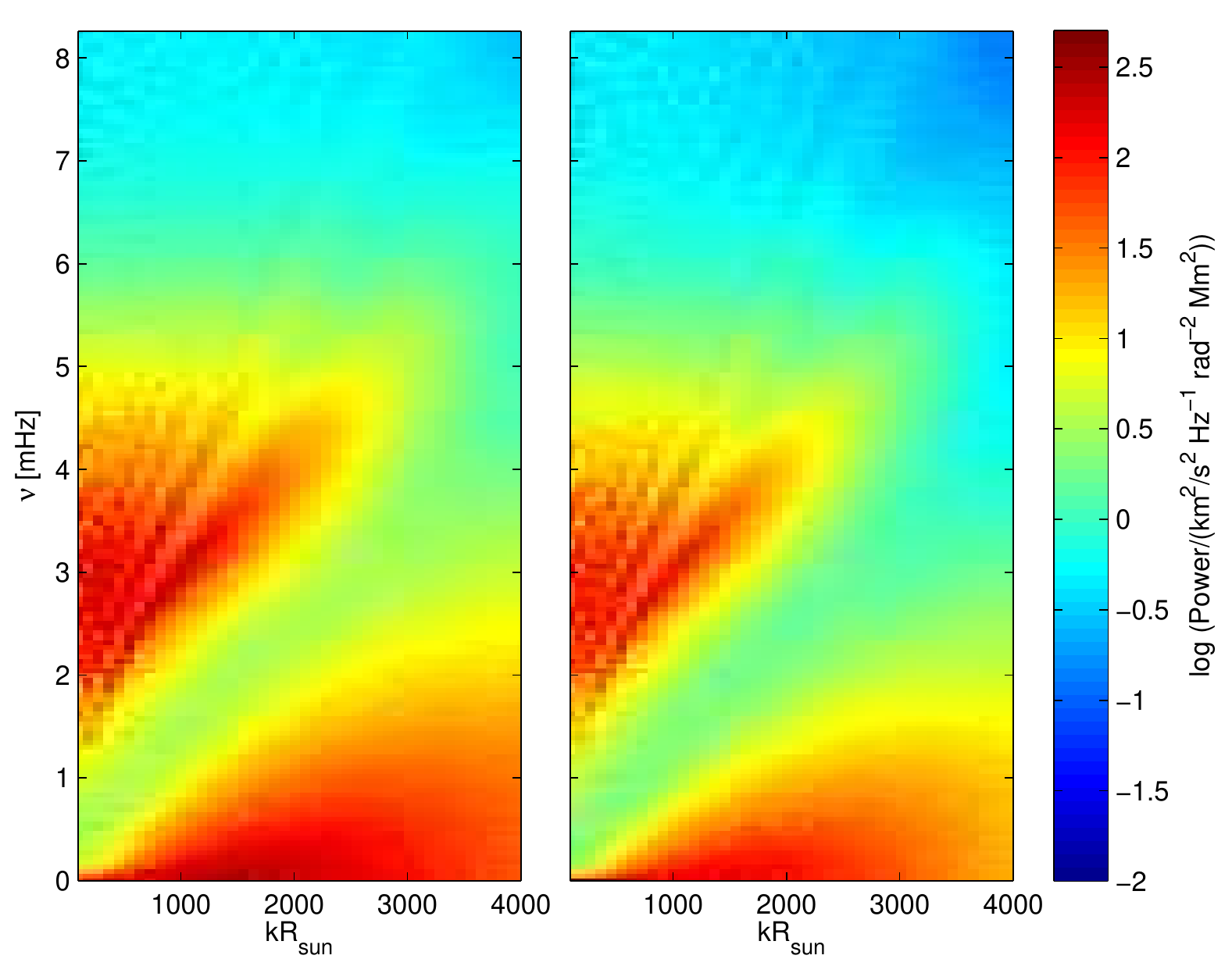}
\caption{Azimuthal average of two power spectra at disk center. {\it Left}: Power for velocities determined directly from the line profiles, {\it right}: Power for velocities computed from simulated instrumental data after running SOPHISM. Both images reveal the solar oscillations and granulation. However, the simulated instrument decreases the power, especially for high wavenumbers. This is caused mostly by the Modulation Transfer Function. Here we use an MTF corresponding to a diffraction-limited instrument (the PSF is an Airy function)}
\label{fig:compare_power}
\end{figure}

We have a time-series of Dopplergrams and thus we can evaluate the effect of the PHI instrument on the measurement of solar oscillations. The most basic of helioseismic diagnostics is the power spectrum as a function of horizontal wavenumber, $k$, and frequency, $\nu$. 
In Figure~\ref{fig:compare_power} we compare two power spectra for an observation at disk center: one generated from synthetic PHI data (i.e. after running SOPHISM) and another one without any influence of the instrument (i.e. for velocity maps derived from the SPINOR line profiles). We see that the power resulting from the synthetic PHI data is lower than the power originating from the line profiles at high-$k$ modes. This is caused by the MTF, which we assume here to correspond to a diffraction-limited instrument.

\begin{figure}
\begin{minipage}[h!]{\textwidth}
\includegraphics[width=0.5\textwidth]{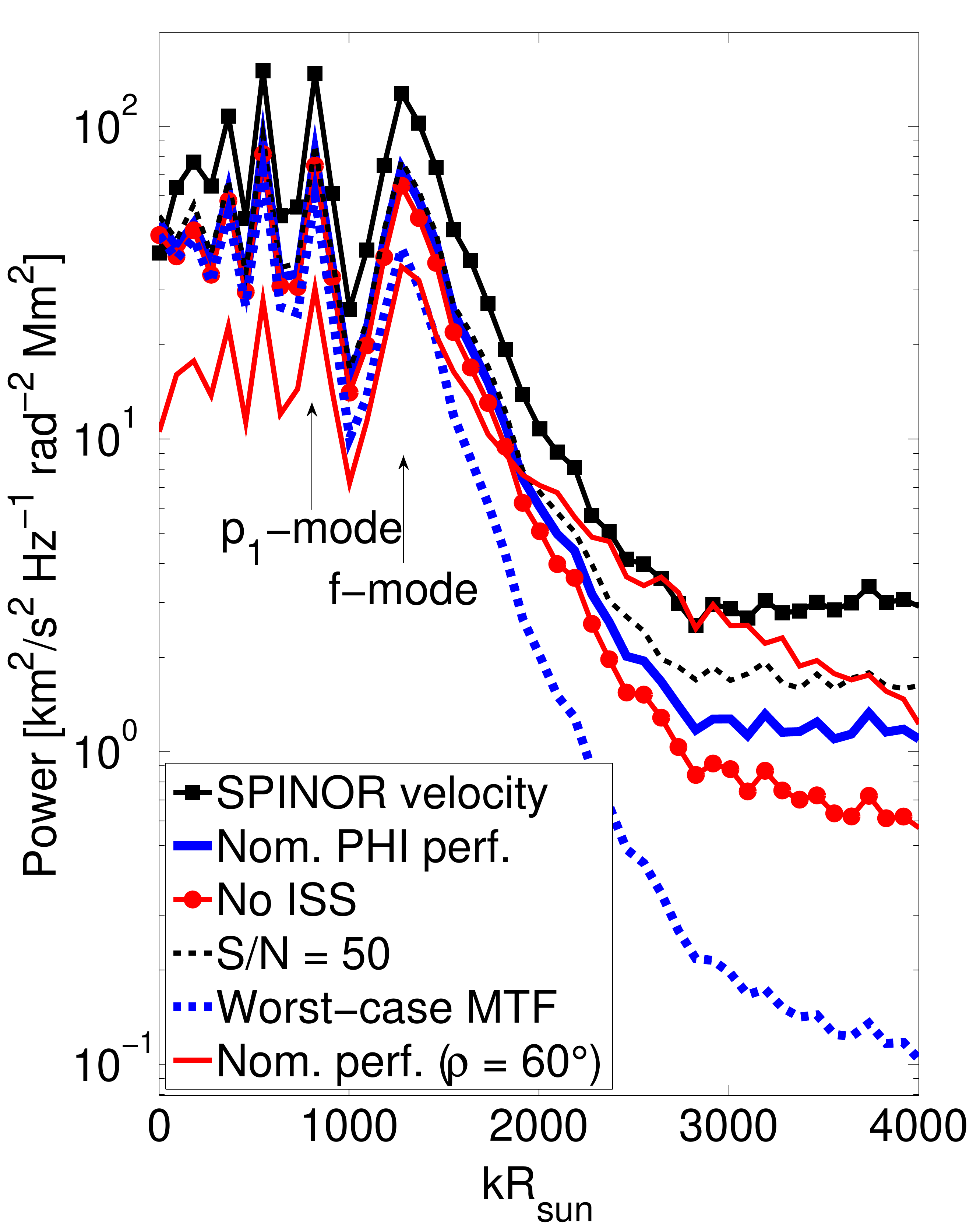}
\includegraphics[width=0.5\textwidth]{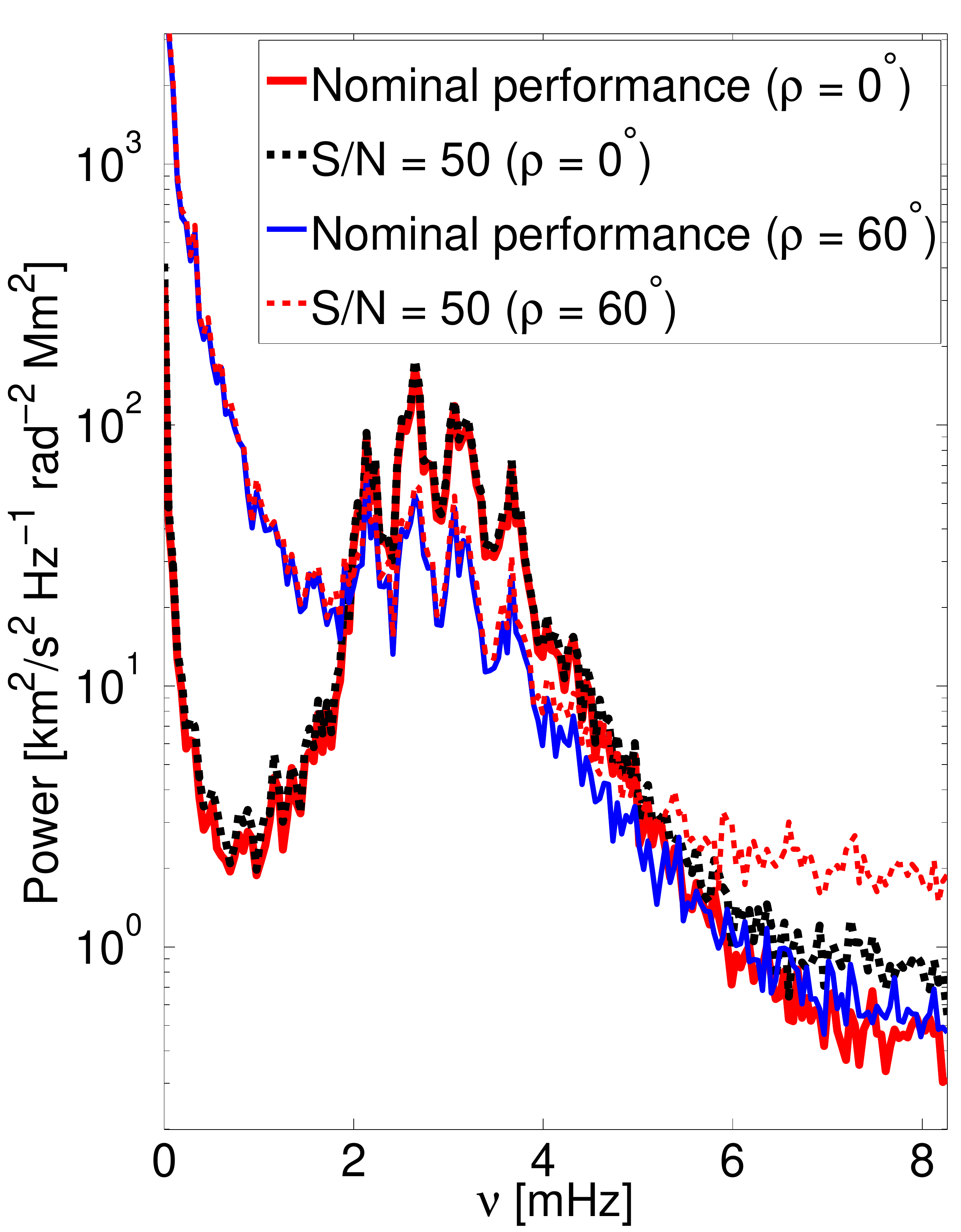}
\end{minipage}
\caption{{\it Left}: Azimuthal average of the power as a function of wavenumber at $\nu=3.41$~mHz. Six configurations are shown, {\it black squares}: Power derived from SPINOR velocities, {\it blue thick solid curve}: Nominal PHI performance, {\it red dots}: ISS turned off, {\it black dashed curve}: Enhanced photon noise (S/N = 50), {\it blue thick dashed curve}: Worst-case MTF, {\it red solid curve}: Nominal PHI performance at $\rho = 60^\circ$. {\it Right}: Power as a function of frequency at $kR_\odot = 456$. {\it Red thick solid line}: Nominal PHI performance when observing at disk center, {\it blue solid line}: Nominal PHI performance at $\rho = 60^\circ$, {\it black thick dashed line}: Enhanced photon noise (S/N = 50) at disk center, {\it red dashed line}: Enhanced photon noise (S/N = 50) at $\rho = 60^\circ$. We do not show the jitter with the ISS off, since it does not depend on frequency. The photon noise enhances the background power. For the nominal PHI performance, the influence of jitter and 
photon noise is negligible. If the ISS is turned off, the high-$k$ power is reduced, a higher photon noise level increases the background power. However, a signal-to-noise ratio (S/N) of 50 is much lower than anything we expect}
\label{fig:power_cuts}
\end{figure}

The impact of the MTF on the power spectrum strongly differs between the two MTFs used in this study (see Sect.~\ref{sect:Optics}). Figure~\ref{fig:power_cuts} shows the azimuthal averages of power spectra for different setups of the instrument. The worst-case MTF reduces the power significantly more at high wavenumbers than the diffraction-limited one. The MTF has a much larger influence on the power spectrum than other instrumental effects, like spacecraft jitter or photon noise. We also show in Figure~\ref{fig:power_cuts}  power spectra, where the ISS is turned off and the photon noise is increased (S/N = 50). Photon noise and jitter only influence the power spectra if they are much worse than the PHI requirements. The photon noise leads to a constant offset in the power spectrum which is negligible compared to the power of the oscillation modes for any noise level that we expect. The spacecraft jitter reduces the power of the high-$k$ modes,
 if the ISS is turned off. All these effects barely affect the oscillations.


When observing at $\rho = 60^\circ$, the power of the p-modes is lower than at disk center, since the displacement caused by the oscillations is mostly radial. On the other hand, the power resulting from convection is increased. Foreshortening leads to a stronger influence of the MTF and spacecraft jitter. Also, the intensity is lower due to limb darkening, which leads to an increased photon noise level.


In short, the most important instrumental effect is the MTF, which attenuates p-mode power at high spatial frequencies. The two MTFs used in this study lead to strong differences in the power spectrum. By comparison, the uncertainties in the jitter or photon noise level have less of an effect. However, we find that oscillations are still easy to observe even with the worst-case MTF.

\section{Lessons Learned from MDI and HMI}\label{sect:lessons}

While the MDI and HMI instruments and associated spacecraft (SOHO and SDO) are in some way quite different from PHI and {\it Solar Orbiter}, the experience from those investigations may nonetheless provide some guidance for PHI.

MDI is in many ways similar to PHI. Telemetry is severely restricted, observables computations are done onboard and the onboard software is complex. A major difference is that {\it Solar Orbiter} has significantly more onboard storage, which may permit post-event data selection.

HMI is significantly different in these respects. The telemetry rate is not a limiting factor, (nearly) raw CCD images are downlinked and the onboard software is quite simple. Having said that, PHI is closer to HMI than to MDI in some respects, including thermal perturbations (due to eclipses), rapid LOS velocity changes (due to the orbit), and the ability to make full polarimetric measurements.

One major lesson learned from MDI is that fragmentation of data has to be controlled. Excessive changes between observing modes make it difficult to find good sequences for other uses than those initially envisioned and make it difficult to search through the data. For helioseismology, it is often particularly important that long sequences are available. In the extreme case of global mode analysis, typical lengths are several months and it is essential that the duty cycle is not too low. Duty cycles from existing helioseismic instruments ranges from 80-90\% for GONG and up to well over 99\% for MDI Medium-$l$ for some 72 day periods. As such we expect that a duty cycle greater than 80\% may be acceptable. However, the amount of degradation will depend on the structure of the gaps. Variable duty cycles can lead to variations in the observed quantities which can in turn be confused with solar variations. But even in the case of local helioseismology, time-series of a significant length are needed and for 
the study of such things as active region evolution it is essential to observe continuously for several days.

Similarly, uncomplicated regular observations from MDI have provided most of the data used for publications. This includes the regular 96 minute cadence magnetograms, the so-called structure program
~\citep[in particular the Medium-$l$ program,][]{1996kosovichev,1997SoPh..170...43K} and the yearly dynamics programs. The Medium-$l$ program, for example, has probably provided the frequencies and splittings used for most publications in global mode seismology. Open and easy access to data have also been essential to ensure the maximum use of the data.

Well calibrated data are also highly desirable. MDI was not particularly well calibrated from the outset, which combined with the onboard observable calculation led to some issues, especially regarding the magnetograms~\citep{2004SoPh..219...39L,2007SoPh..241..185L,2012SoPh..279..295L}. Another problem for MDI has been that the noise in the shutter~\citep{2004SoPh..219...39L} also leads to an error in the velocity zero point, which in turn introduces a lot of noise in the low degree modes. HMI, by comparison, was substantially better calibrated from the start~\citep{2012SoPh..275..285C,2012SoPh..275..261W,2012SoPh..275..327S}.

The issue of calibrations is particularly difficult for PHI, due to the need to do onboard observables calculations and the massive thermal perturbations. For HMI all the raw data are available on the ground, which means that computations can be done and, if necessary, redone. For PHI this will not be the case and it is thus essential that enough data are downlinked early in the mission to ensure that good calibrations can be derived and adequate
onboard algorithms can be developed and tested before regular observations start. It will also be necessary to find some way to ensure that any drifts in the calibrations will be detected and corrected.

Having said that, some of these programs were not designed as well as they could have been, especially in hindsight. In particular, some of the data products could probably have been compressed more (e.g. by truncating more of the bottom bits) and a better trade-off between truncation noise, smoothing, subsampling density and crop radius could probably have been done for the Medium-$l$ program. It might have been better to truncate more bits while keeping a larger FOV. Given the desire to have long and uniform time-series the trade-off was not adjusted during the mission.

A possibility with {\it Solar Orbiter} is to take data, store it on board and decide later if it should be downlinked. This may be particularly interesting for emerging active regions, flares, and so forth. There is obviously a trade-off involved here, as this will increase operational complexity and fragmentation of the data.



\begin{acknowledgements}
This work is supported by the European Commission through FP7 project ``Exploitation of Space Data for Innovative Helio- and Asteroseismology'' (SpaceInn). We are very grateful to R.~Stein for making the Stagger simulation data available. We thank A.~Lagg for his support with using the SPINOR code. B.L. is a member of the International Max Planck Research School for Solar System Science at the University of G\"ottingen. B.L. designed and performed the research described in section~\ref{sect:synth_data} and wrote this section. B.L. performed research for section~\ref{sect:science} and contributed to the writing of sections \ref{sect:intro},\ref{sect:mission},\ref{sect:science} and \ref{sect:PHI}.
\end{acknowledgements}

\bibliographystyle{aps-nameyear}

\bibliography{ISSI_Solar_Orbiter}

\nocite{*}

\end{document}